\documentclass[aps,pra, twocolumn,a4paper,showpacs]{revtex4}
\usepackage{bm}
\usepackage{amssymb,amsmath,amsfonts,graphicx}

\newcommand{\ddpt}{\frac{\partial}{\partial t}}

\newcommand{\Hs}{\mathbb{H}}

\newcommand{\Cs}{\mathbb{C}^2}
\newcommand{\un}{\openone}

\def\BE {\begin{equation}}
\def\EE {\end{equation}}
\def\BEA {\begin{eqnarray}}
\def\EEA {\end{eqnarray}}
\def\BES {\begin{subequations}}
\def\EES {\end{subequations}}
\def\BA {\begin{array}}
\def\EA {\end{array}}
\def\NN {\nonumber}
\def\ep {\epsilon}
\def\epnm {\epsilon^{2^{n-1}}}
\def\epn {\epsilon^{2^n}}

\def\epkm {\epsilon^{2^{k-1}}}

\def\ad{{\rm ad}}
\def\Mag{M_{H_1}}

\def\ef{^{{\rm e}}}
\def\efz{^{{\rm e}}_0}
\def\efu{^{{\rm e}}_1}
\def\efd{^{{\rm e}}_2}

\def\efk{^{{\rm e}}_k}
\def\efkm{^{{\rm e}}_{k-1}}

\def\efn{^{{\rm e}}_{n}}

\def\inte{^{{\rm i}}_1}
\def\intz{^{{\rm i}}_0}
\def\intu{^{{\rm i}}_1}

\def\eintn{^{{\rm e i}}_n}
\def\eintk{^{{\rm e i}}_k}

{

\begin{document}

\title{Pulse-driven quantum dynamics beyond the impulsive regime}
\author{D.~Daems}
\email{ddaems@ulb.ac.be}
\affiliation{Center for Nonlinear Phenomena and Complex Systems, Universit\'e Libre de
Bruxelles, CP 231, 1050 Brussels, Belgium}
\author{S.~Gu\'erin }
\email{sguerin@u-bourgogne.fr}
\author{H. R.~Jauslin}
\affiliation{Laboratoire de Physique de l'Universit\'e de Bourgogne, UMR CNRS 5027, BP
47870, 21078 Dijon, France}
\author{A.~Keller}
\email{arne.keller@ppm.u-psud.fr}
\author{O.~Atabek}
\affiliation{Laboratoire de Photophysique Mol\'eculaire du CNRS, Universit\'e Paris-Sud,
B\^at. 210 - Campus d'Orsay, 91405 Orsay Cedex, France}

\begin{abstract}
We review various unitary time-dependent perturbation theories and
compare them formally and numerically. We
show that the Kolmogorov-Arnold-Moser technique performs better owing to both the superexponential character of correction terms and the possibility to optimize the accuracy of a given level of approximation which is explored in details here.
As an illustration, we consider a two-level system driven by short pulses beyond the sudden limit.
\end{abstract}

\maketitle


\section{Introduction}

Short and intense laser pulses allow nowadays to drive atoms and molecules
in nonperturbative regimes going from adiabatic (nano- and picosecond) to
sudden or impulsive (femtosecond). Recent examples concern the alignment of
molecules, which can be achieved during nanosecond pulses or after
femtosecond pulses \cite{align}. Corrections to perfect adiabaticity can be
analyzed in terms of superadiabatic \cite{Berry90,Berry93,Joye93,Holthaus}
and Davis-Dykhne-Pechukas techniques \cite{Dykhne,Davis,Joye91}. On the
opposite side, regimes beyond the impulsive approximation, i.e. beyond the
limiting case of pulses described as $\delta -$kicks, have not been yet much
explored due to a lack of adapted tools of analysis.

It is well-known that one can treat periodic perturbations using extended
Hilbert spaces where time is considered as a new dynamical variable, in
order to render the problem autonomous \cite{Sambe, Howland}. This approach,
that can be formulated as Floquet theory \cite{Shirley, Barata,Adv},
allows one to eliminate systematically secular terms (i.e. terms that grow arbitrary with time), which would otherwise lead
to divergences.
Pulse-driven dynamics associated to Hamiltonians localized in time requires
a different treatment of secular terms. In this case, since the perturbation
acts only during a finite time interval, the secular terms do not lead to
divergences.

This article contributes to develop a time-dependent perturbation technique, that is in particular suited for pulse-driven dynamics, on the basis of Refs. \cite{art1,RC}. 
In Ref. \cite{art1}, we constructed a superexponential perturbation theory which preserves the unitarity of the evolution
operator at each order, and applied it beyond the impulsive
regime by considering an expansion where the perturbative parameter is the characteristic
duration of the time-dependent interaction compared to the characteristic
time for the free evolution.
We have shown that it converges in any regime (from impulsive to adiabatic) in two-level
systems. This derivation is based on the Kolmogorov-Arnold-Moser (KAM)
technique applied in an extended Hilbert space \cite{Bellissard,Combescure,Blekher,Duclos,SchererIII}. 
In Ref. \cite{RC}, we presented an improvment of this technique taking advantage of
free parameters, connected to secular terms, that are  available to reduce the error
without prior knowledge of the exact solution. This optimization enhances the
accuracy of the method in such a way that the first order approximation
gives a satisfactory description up to fairly large values of the perturbative parameter. 

The present article contains a detailed
description of the methods announced in \cite{RC}.
Instead of using an extended space, we formulate the
derivation in a simpler way, by stating the perturbation iterations directly
at the level of the evolution operator in the original Hilbert space.
This scheme allows us to consider and compare in a unified way various time-dependent perturbation techniques. 
In particular we make the connection with the well-known Magnus
expansion \cite{Magnus}, that has been used by Henriksen {\em et al.} to
construct an improved impulsive approximation \cite{Henriksen}. 
We also develop and investigate the accuracy optimization which can be applied to the time-dependent Poincar\'e-Von Zeipel, the time-dependent Van Vleck and the time-dependent KAM techniques.

The paper is organized as follows. In Sec.~\ref{sec:time} we recall the Magnus expansion and outline the time-dependent versions of the Poincar\'e-Von Zeipel, the Van Vleck and  the KAM techniques.
We highlight the free parameters and free operators that may be present in these unitary  perturbative methods.
In Sec.~\ref{sec:accuracy} we exploit these degrees of freedom to improve  the accuracy of a given level of approximation.
Section \ref{sec:sudden} is devoted to the application of these techniques beyond the impulsive regime and the illustration on a pulse-driven two-level system.
The conclusions are given in Sec.~\ref{sec:conclusion} and some details of the calculations are reported in Appendixes A to C.

\section{Unitary time-dependent perturbation theories}
\label{sec:time}
We consider the Schr\"odinger equation
\BES
\label{eq:H1UH1}
\BEA
\label{eq:Xdot}
i\ddpt U_{H_1}(t,t_0) =H_1(t)\,
U_{H_1}(t,t_0), \quad U_{H_1}(t_0,t_0)=\un , \quad
\EEA
where $H_1(t)$ is a time-dependent matrix or a time-dependent operator in a Hilbert space $\Hs$. 
Assuming that one can decompose $H_1(t)$ according to
\BEA
\label{eq:M0M1}
H_1(t)=H_0(t)+\ep V_1(t) ,
\EEA
\EES
where $H_0(t)$ is such that its propagator $U_{H_0}(t,t_0)$ is known  and $V_1(t)$ is localized in time (i.e., vanishes outside a finite interval), we are looking for a unitary perturbative expansion of the full propagator $U_{H_1}(t,t_0)$.
This can be achieved by two classes of techniques that we outline below:
i) the order by order methods, namely the Magnus expansion, the time-dependent Poincar\'e-Von Zeipel technique and the time-dependent  Van
Vleck technique, where after $n$ steps the remainder is of order $\ep^{n+1}$; and
ii) the surperexponential KAM technique where the remainder is of order $\epnm$.

Below we will have to consider the propagator $U_{H_1}(t,t_0)$ in the interaction representation with respect to $H_0(t)$
\BES
\label{eq:interaction}
\BEA
 \label{eq:UHIMag}
 U_{H\inte}(t,t_0;s)\equiv   U_{H_0}(s,t) U_{H_1}(t,t_0)U_{H_0}(t_0,s),
 \EEA
 where $s$ is an arbitrary time (the standard interaction representation corresponds to the case $s=t_0$).
 This propagator satisfies the Schr\"odinger equation
\BEA
\label{eq:XIdot}
i\ddpt U_{H\inte}(t,t_0;s) =H\inte(t;s)\,
U_{H\inte}(t,t_0;s) ,\qquad
\EEA
and the associated Hamiltonian reads
\BEA
 H\inte(t;s) \equiv  \ep \,U_{H_0}(s,t)  V_1(t)U_{H_0}(t,s) .\qquad
 \label{eq:HIMag}
\EEA
\EES
We will also consider a new representation defined with the help of a unitary transformation $T(t;s)$ according to
\BEA
 \label{eq:T}
 U_{H}(t,t_0;s)\equiv   T^\dagger(t;s) U_{H_1}(t,t_0)T(t_0,s),
 \EEA
 where $H(t;s)$ is a new Hamiltonian.
This expression is reminiscent of Eq.~(\ref{eq:UHIMag}) although $T(t;s)$ need not be a propagator but a unitary operator which features an arbitrary parameter $s$ and satisfies the property $T^\dagger(t;s)=T(s;t)$.

\subsection{Magnus expansion}
\label{sec:Mag}
The solution to Eq.~(\ref{eq:Xdot}) can always be put in  the form of an exponential
\BEA
\label{eq:X}
U_{H_1}(t,t_0)=e^{-i\Mag(t;t_0)},\quad \Mag(t_0;t_0)=0 ,
\EEA
where $\Mag(t;t_0)$ is self-adjoint to ensure unitarity.
The exponent $\Mag(t;t_0)$ is generally not simply the integral of $H_1(t)$ owing to the non-commutativity of this latter for different times.
Indeed from Eq.~(\ref{eq:Xdot}) one deduces the following equation for $\Mag(t;t_0)$
\begin{widetext}
\BEA
\label{eq:Magdot}
\ddpt \Mag(t;t_0)=H_1(t)+\frac{i}{2}\left[\Mag(t;t_0),H_1(t)\right]-\frac{1}{12}\left[\Mag(t;t_0),\left[\Mag(t;t_0),H_1(t)\right]\right]+ \cdots .\quad
\EEA
We refer to the paper of Magnus \cite{Magnus} for a derivation of this equation (see also Ref. \cite{Pechukas}).
The matrix or the operator $\Mag(t;t_0)$ is obtained by integrating  the first term on the right hand side of Eq.~(\ref{eq:Magdot}) and substituting the result into the next terms of this equation.
One then  repeats  this procedure, known as Picard's iteration, with the resulting second and subsequent terms.
This   yields the Magnus expansion
\BEA
\label{eq:Mag}
\Mag(t;t_0)\!&\!=\!&\!\int_{t_0}^t d u H_1(u)+\frac{i}{2}\int_{t_0}^t d u \left[\int_{t_0}^u d v H_1(v),H_1(u)\right]-\frac{1}{4}\int_{t_0}^t d u \left[\int_{t_0}^u d v \left[\int_{t_0}^v d w H_1(w),H_1(v)\right],H_1(u)\right]\NN\\
\!&\!-\!&\!\frac{1}{12}\int_{t_0}^t d u \left[\int_{t_0}^u d v H_1(v),\left[ \int_{t_0}^u d w H_1(w),H_1(u)\right]\right]+\cdots .
\EEA
\end{widetext}
The number of terms in this expansion grows very rapidly \cite{Iserles}.
Notice that this expansion is not limited to perturbation theory [ i.e. to a Hamiltonian of the form of Eq.~(\ref{eq:M0M1})] although it is of interest only when the subsequent terms are negligible.
This will be the case if one is interested in obtaining an expression valid for very short times only or in the presence of another small parameter.

In the framework of perturbation theory, if we were to substitute Eq.~(\ref{eq:M0M1}) into Eq.~(\ref{eq:Mag}) we would generally obtain contributions to a given order in $\ep$ from an infinite series of terms of the Magnus expansion \footnote{Notice that all the terms containing solely $H_0(t)$ are also given by a series as that of Eq. (\ref{eq:Mag}) where $H_1(t)$ is replaced by $H_0(t)$, and yield, as a consequence, the propagator $U_{H_0}(t,t_0)$. By going to the interaction representation we get rid of this particular series, and factor out this propagator right from the start.}.
This can easily be circumvented by considering Eq.~(\ref{eq:interaction}), the equivalent of Eq.~(\ref{eq:H1UH1}) in the interaction representation. 
The Magnus expansion pertaining to Eq.~(\ref{eq:XIdot}) allows one to write its solution in the form
\BEA
\label{eq:UHiM}
 U_{H\inte}(t,t_0;s)=\exp \left\{-i\ep M_{H\inte}(t;t_0,s)\right\} ,
\EEA
with $\ep M_{H\inte}(t;t_0,s)$ given by Eq.~(\ref{eq:Mag}) where $H_1(t)$ is replaced by $H\inte(u;s)$. 
By virtue of Eq.~(\ref{eq:HIMag}), each $H\inte(u;s)$ carries a prefactor $\ep$.
Hence, there is no $\ep$-independent term and terms of $M_{H\inte}(t;t_0,s)$ with higher numbers of  $H\inte(u;s)$ are now of higher orders in $\epsilon$.
In order to display the explicit dependence on the parameter $s$, we decompose the propagator entering through $H\inte(u;s)$ according to $U_{H_0}(s,u)=U_{H_0}(s,t)U_{H_0}(t,u)$ and  note that the leftmost $U_{H_0}(s,t)$ and rightmost $U_{H_0}(t,s)$ of each term of $M_{H\inte}(t;t_0,s)$ factor out (while the inner ones cancel each other)
\BEA
\label{eq:MHi1}
\ep M_{H\inte}(t;t_0,s)=U_{H_0}(s,t)\sum_{k=1}^{\infty} \ep^k M_k(t;t_0)U_{H_0}(t,s). \quad
\EEA
Returning to the original representation with the help of Eq.~(\ref{eq:UHIMag}), Eq.~(\ref{eq:UHiM}) yields
\BEA
 U_{H_1}(t,t_0)= \exp\left\{-i\sum_{k=1}^{\infty} \ep^k M_k(t;t_0)\right\} U_{H_0}(t,t_0),\qquad\label{eq:UHMagbis}
\EEA
where use was made of the 
identity 
\BEA
\label{eq:idexp}
A e^B A^{-1}=e^{A B A^{-1}}.
\EEA

By truncating  the infinite series of Eq.~(\ref{eq:UHMagbis}) to order $n$ one obtains the unitary approximation 
\BE
\label{eq:UH1UHnMag}
U_{H_1}(t,t_0)=U_{H_1}^{(n)}(t,t_0)+\mathcal{O}(\ep^{n+1}),
\EE
with the Magnus expansion
\BEA
\label{eq:UH1nMag}
 U_{H_1}^{(n)}(t,t_0)=\exp\left\{-i\sum_{k=1}^{n} \ep^k M_k(t;t_0)\right\} U_{H_0}(t,t_0). \qquad
\EEA
We stress that this expression is independent of the arbitrary time  $s$ chosen in Eq.~(\ref{eq:interaction}) \footnote{In Ref.~\cite{Henriksen} a time parameter is introduced through an interaction representation and shown to affect the final result. This is due to the fact that in Ref.~\cite{Henriksen} an additional approximation is made which assumes an impulsive character for the time-dependent perturbation.}.
Each term of the exponential can be cast into the form
\BEA
\label{eq:MnMag}
 M_k(t;t_0)=\int_{t_0}^t \!d u  U_{H_0}(t,u) V_k(u)U_{H_0}(u,t),
\EEA
where $V_k(u)$ is deduced from Eqs.~(\ref{eq:HIMag}),~(\ref{eq:Mag}) and (\ref{eq:MHi1}).
The case $k=1$ features the perturbation itself while for the first few values of $k$  that we shall use below, dropping the time arguments, one finds
\BES
\BEA
\label{eq:VnMag}
  V_2 \!\!&=& \!\!
 -\frac{i}{2}\left[M_1,V_1\right],\\
  V_3 \!\!&=& \!\!
 -i\left[M_2,V_1\right] +\frac{1}{6}
\left[M_1,\left[M_1,V_1\right]\right] . \qquad
\EEA
\EES

\subsection{Time-dependent Poincar\'e-Von Zeipel expansion}
\label{sec:PV}
The classical Poincar\'e-Von Zeipel technique was adapted to quantum mechanics by Scherer  to treat both time-independent \cite{SchererI} and time-dependent \cite{SchererIII} sytems.
In the time-independent case it was shown \cite{SchererI} that this technique coincides with the usual Rayleigh-Schr\"odinger expansion.
We present in Appendix \ref{app:PV} a time-dependent Poincar\'e-Von Zeipel technique that has the advantages to be strictly unitary upon truncation at any given order and does not require the consideration of a so-called extended Hilbert space.
This method amounts to map the full propagator $U_{H_1}(t,t_0)$ into a new effective propagator $U_{H\ef}(t,t_0)$ with the help of a single unitary transformation according to an equation similar to Eq.~(\ref{eq:T}).
Moreover, this formulation exhibits free parameters available to improve the accuracy of a given step of the algorithm by a procedure we describe in Sec. \ref{sec:optim}.
Finally, we show that the time-dependent Poincar\'e-Von Zeipel method includes the Magnus expansion as a particular case.

\subsection{Time-dependent Van Vleck technique}
\label{sec:VV}
To provide a general perspective of unitary time-dependent perturbation theories we introduce in Appendix \ref{app:VV} a time-dependent version of the Van Vleck technique that is more widely used than the preceding method in the stationary case \cite{VanVleck}.
It consists in transforming the full propagator into a new effective propagator iteratively through a series of unitary operations.
Our purpose is actually  not to introduce yet another variant of perturbation theory but to emphasize that this time-dependent version of a well-known technique is i) comparable to the Poincar\'e-Von Zeipel method which we show to be closely related to the Magnus expansion, and ii) not as performant as the KAM technique detailed below since it is an order by order perturbative method.

\subsection{Time-dependent Kolmogorov-Arnold-Moser expansion}
\label{sec:KAM}
The time-dependent KAM technique aims at obtaining 
a superexponential expansion for the propagator $U_{H_1}(t,t_0)$
through a series of unitary transformations.
It is sometimes said to be superconvergent.
However, superexponential is more appropriate  since the dependence  on $\ep$ of the remainder after $n$ iterations is indeed  the exponential of an exponential ($\epnm$)  while the actual convergence of the algorithm has to be examined specifically.
\subsubsection{First iteration}
\label{sec:KAM1}
The first step is to construct a unitary operator $T_1(t)$ which  transforms the propagator $U_{H_1}(t,t_{0})$ we are looking for into the propagator $U_{H_2}(t,t_{0})$ [cf. Eq.~(\ref{eq:T})]
\BEA
\label{eq:UH1UH2KAM}
T_1^\dagger(t)U_{H_1}(t,t_0)T_1(t_0)=U_{H_2}(t,t_0),\quad
\EEA
where  $U_{H_2}(t,t_0)$ is associated with the sum of an  effective Hamiltonian
$H\efu(t)$ which contains all contributions up to order $\ep$ and  a remainder $\ep^2 V_2(t)$ 
\BEA
\label{eq:H2KAM}
H_2(t)\equiv  H\efu(t) + \ep^2 V_2(t).
\EEA 

As the new propagator is  generated by a sum of Hamiltonians
it can be expressed in terms of the propagator $U_{H\efu}(t,t_0)$   and a propagator $U_{R_2}(t,t_{0})$ by considering the interaction representation with respect to $H\efu(t;t_1)$
\BEA
\label{eq:UH2KAM}
U_{H_2}(t,t_0)=U_{H\efu}(t,t_1^{\prime\prime})U_{R_2}(t,t_{0};t_1^{\prime\prime})
U_{H\efu}(t_1^{\prime\prime},t_0).
\EEA
By virtue of Eq.~(\ref{eq:interaction}) the propagator $U_{R_2}(t,t_0;t_1^{\prime\prime})$ satisfies a Schr\"odinger equation whose Hamiltonian is 
\BEA
\label{eq:R2KAM}
\ep^2 R_2(t;t_1^{\prime\prime}) \equiv \ep^2 U_{H\efu}(t_1^{\prime\prime},t) V_2(t)
U_{H\efu}(t,t_1^{\prime\prime}) .
\EEA
Being associated to a Hamiltonian of second order in $\ep$, this propagator will be neglected in Eq.~(\ref{eq:UH2KAM}), i.e., replaced by the identity.

The effective Hamiltonian can be decomposed as
\BEA
\label{eq:He1KAM}
H\efu(t)\equiv  H_0(t) + \ep D_1(t),
\EEA
which allows one to express $U_{H\efu}(t,t_0)$ as
\BEA
\label{eq:UHe1KAM}
U_{H\efu}(t,t_0)=U_{H_0}(t,t_1)U_{P_1}(t,t_0;t_1)U_{H_0}(t_1,t_0),
\EEA
where $U_{P_1}(t,t_0)$ is the propagator corresponding to  the Hamiltonian 
\BE
\ep P_1(t;t_1)\equiv \ep U_{H_0}(t_1,t)D_1(t)U_{H_0}(t,t_1).
\EE
Thus far the only restriction on the Hamiltonian $\ep D_1(t)$ is that it be of order $\ep$.
Hence we have the freedom to choose it so as to be able to determine explicitly the propagator $U_{P_1}(t,t_0;t_1)$.
This will be the case if 
\BEA
\label{eq:D1KAM}
D_1(t) \equiv U_{H_0}(t,t_1)D_1(t_1;t_1)U_{H_0}(t_1,t) \equiv  D_1(t;t_1), \quad
\EEA
with $D_1(t_1;t_1)$ arbitrary.
In this paper, we consider explicitly two possibilities
\BES
\label{eq:D1sKAM}
\BEA
\label{eq:D10KAM}
D_1(t_1;t_1)\!&\!\equiv \!&\!0,\quad \\
\label{eq:D1V1KAM}
D_1(t_1;t_1)\!&\!\equiv \!&\!V_1(t_1).\quad
\EEA
\EES
The first one is a trivial choice which gives  nevertheless a nontrivial one-iteration KAM expansion (it will be shown in Sec.~\ref{sec:comp} to coincide, for the first iteration only, with the first order Magnus expansion).
The choice to relate $D_1(t_1;t_1)$ to the perturbation  according to Eq.~(\ref{eq:D1V1KAM}) is also rather natural as this operator enters the effective Hamiltonian (we shall discuss the fact that the perturbation is evaluated at an arbitrary time $t_1$ in Sec.~\ref{sec:limit}).
Choosing one of the possibilities of Eq.~(\ref{eq:D1sKAM}) implies that the propagator defined in Eq.~(\ref{eq:UHe1KAM}) reads
\BEA
\label{eq:S1KAM}
U_{P_1}(t,t_0;t_1)= e^{-i (t-t_0) \ep D_1(t_1;t_1)}.
\EEA
The effective propagator can then be given a convenient form using Eqs.~(\ref{eq:idexp}), (\ref{eq:UHe1KAM}) and (\ref{eq:D1KAM})
\BEA
U_{H\efu}(t,t_0)\!&\!=\!&\!
U_{H_0}(t,t_0)  e^{-i (t-t_0) \ep D_{1}(t_0;t_1)}\NN\\ \!&\! \equiv \!&\!  U_{H\efu}(t,t_0;t_1) .\quad
\label{eq:UHe1D1KAM}
\EEA

We now express the requirement that $T_1(t)$ defined in Eq.~(\ref{eq:UH1UH2KAM}) be such that $\ep^2 V_2(t)$ contain no terms of order lower than $\ep^2$.
We first multiply Eq.~(\ref{eq:Xdot}) from the left by $T_1^\dagger(t)$ and from the right by $T_1(t_0)$ to deduce employing also Eq.~(\ref{eq:UH1UH2KAM})
\BEA
\label{eq:coho1KAM}
\ep^2 V_2(t)=T_1^\dagger(t)H_1(t)T_1(t)- H\efu(t)-T_1^\dagger(t) i\frac{\partial}{\partial t}T_1(t).\quad
\EEA
Writing  $T_1(t)$  in the exponential form
\BEA
\label{eq:T1KAM}
T_1(t)\equiv e^{-i\ep W_1(t)} , 
\EEA
we then require that all terms of order lower than $\ep^2$ in Eq.~(\ref{eq:coho1KAM}) vanish identically.
This leads to a  differential equation for  the self-adjoint operator $W_1(t)$
\BE
\label{eq:W1dotKAM}
\frac{\partial}{\partial t}
W_1(t)=V_1(t)-D_1(t;t_1)+i\left[W_1(t),H_0(t)\right] ,
\EE
and defines the remainder $\ep^2 V_2(t)$ as the right side of  Eq.~(\ref{eq:coho1KAM}).
We stress that to arrive at Eq.~(\ref{eq:W1dotKAM}) we do not identify terms order by order.
Hence this equation is still valid if any of the operator featured  contains a further dependence on $\ep$.
The  general solution to Eq.~(\ref{eq:W1dotKAM}) reads
\BEA
\label{eq:W1KAM}
W_1(t)\!&\!=\!&\!\int_{t_1^\prime}^t d u U_{H_0}(t,u)\left[V_1(u)-D_1(u;t_1)\right]U_{H_0}(u,t) \NN\\ 
\!&\!+\!&\!U_{H_0}(t,t_0)B_1 U_{H_0}(t_0,t),\quad
\EEA
where $B_1$ is any constant self-adjoint operator.
In the present work we choose $B_1=0$.

Substituting Eq.~(\ref{eq:UH2KAM})  into Eq.~(\ref{eq:UH1UH2KAM}) and replacing $U_{R_2}(t,t_0;t_1^{\prime\prime})$ by the identity as discussed above, one obtains the KAM approximation
\BE
\label{eq:UH1UH11KAM}
U_{H_1}(t,t_0)=U_{H_1}^{(1)}(t,t_0)+\mathcal{O}(\ep^2) , \qquad
\EE
with the one-iteration KAM expansion
\BEA
\label{eq:UH11KAM}
U_{H_1}^{(1)}(t,t_0)=  {T_1}(t)U_{H\efu}(t,t_0)T_1^{\dagger}(t_0).
\EEA
We emphasize the  dependence of the following operators on
the arbitrary times $t_1$, $t_1^\prime$ and $t_1^{\prime\prime}$
\BEA
H\efu(t)\!&\!\equiv \!&\!H\efu(t;t_1),\NN\\
T_1(t)\!&\!\equiv \!&\!T_1(t;t_1,t_1^\prime),\NN\\
W_1(t)\!&\!\equiv \!&\!W_1(t;t_1,t_1^\prime),\NN\\
V_2(t)\!&\!\equiv \!&\!V_2(t;t_1,t_1^\prime),\NN\\
R_2(t;t_1^{\prime\prime})\!&\!\equiv \!&\!R_2(t;t_1,t_1^\prime,t_1^{\prime\prime}). 
\label{eq:dep1}
\EEA
As a consequence, the one-iteration KAM expansion depends on  $t_1$ and $t_1^\prime$. 
In Sec.~\ref{sec:optim}, we shall show how these parameters can be chosen to improve the accuracy of the algorithm.
In addition, recall that there are two constant operators, 
$D_1(t_1;t_1)$ in Eq.~(\ref{eq:D1KAM}) and $B_1$ in Eq.~(\ref{eq:W1KAM}), that can be freely chosen.
Note that with the choice of Eq.~(\ref{eq:D10KAM}) there is no dependence on $t_1$.

\subsubsection{Second iteration}
In the first iteration of the KAM algorithm we started with 
  the Hamiltonian $H_1(t)= H_0(t) + \ep V_1(t)$ and the known propagator $U_{H_0}(t,t_{0})$.
We constructed, with the help of
$T_1(t)$, a new Hamiltonian $H_2(t)=  H\efu(t) + \ep^2 V_2(t)$ and its propagator $U_{H_2}(t,t_{0})$.
We approximated these operators to first order by retaining only the effective Hamiltonian $H\efu(t)$ and its propagator $U_{H\efu}(t,t_{0})$, discarding thus the remainder $\ep^2 V_2(t)$ and the related propagator $U_{R_2}(t,t_{0};t_1^{\prime\prime})$.

To go one step further, 
unlike standard perturbation theory which reduces the size of the remainder from $\ep^2$ to $\ep^3$, 
the KAM algorithm takes advantage of the fact that after one iteration it produces a new perturbation $\ep^2 V_2(t)$ whose order is the square of that of the original perturbation $\ep V_1(t)$.
Hence by considering $H_2(t)$ and in particular the perturbation $\ep^2 V_2(t)$ as the new starting point, a KAM transformation $T_2(t)$ produces a new perturbation  $\ep^4 V_3(t)$ [whose order is indeed $(\ep^2)^2$].
This is the essence of the superexponential   character of the KAM algorithm.
The fact that the new perturbation is of order $\ep^4$ instead of  $\ep^3$ as in a standard perturbation theory, allows one to anticipate the importance of keeping higher order terms in $\ep^2 V_2(t)$, in particular terms of order $\ep^3$ which would otherwise be absent if one further iterates the algorithm.

The second KAM iteration amounts thus to reproduce the first iteration on the newly constructed Hamiltonian $H_2(t)=  H\efu(t) + \ep^2 V_2(t)$:
 the effective Hamiltonian $H\efu(t)$ and propagator $U_{H\efu}(t,t_0)$ now play the role of the previous unperturbed Hamiltonian and propagator respectively,
we replace $\ep$ by $\ep^2$ and 
 increase each subscript by one unit. 
The perturbation $\ep^2 V_2(t)$ is given by  Eq.~(\ref{eq:coho1KAM}) which we expand using Eqs.~(\ref{eq:T1KAM})-(\ref{eq:W1dotKAM}) and the Hausdorff formula \cite{Hausdorff,Primas}
\BEA
e^A B e^{-A}\!\!&=&\!\!B+\frac{1}{1!}[A,B]+\frac{1}{2!}[A,[A,B]]\NN\\
\!\!&+&\!\!\frac{1}{3!}[A,[A,[A,B]]]+\cdots,
\EEA
to obtain 
\BEA
\ep^2 V_{2}\!\!&=&\!\!\frac{i\ep^2}{2} \left[W_1,V_1+D_1\right]-\frac{\ep^3}{6} \left[W_1,\left[W_1,2V_1+D_1\right] \right]\NN\\
\!\!&-&\!\!
\frac{i\ep^4}{24}\left[W_1, \left[W_1,\left[W_1,3V_1+D_1\right] \right]\right]+\cdots. \qquad \label{eq:V2KAM}
\EEA
We can rewrite this expression in a compact form with  the shorthand notation $\ad^{k}(A,B)$ for the $k$ nested commutators $\left[A,\cdots \left[A,\left[A,B\right]\right] \cdots  \right]$
\BEA
\ad^{k}(A,B)\equiv \left\{
\BA{ll} 
B & k=0\\
\left[A,\ad^{k-1} (A,B) \right] \quad&  k \geq 1  .
\EA 
\right. 
\label{eq:ad}
\EEA
The perturbation we shall start from in the second iteration of the KAM algorithm reads thus
\BE
\label{eq:V2KAMad}
\ep^2 V_{2}=\sum_{k=1}^{\infty}
\frac{i^{k}\ep^{k+1}}{(k+1)!}\,\ad^{k}\left(W_1,kV_1+D_1\right).
\EE
We recall that the presence of a power series in $\ep$ constitutes no difficulty for this algorithm.
It is on the contrary crucial in this superexponential technique to keep terms up to the final order one is interested in.
Terms of order $\ep^3$, for instance, do not appear through $\ep^4 V_3(t)$ but through the second term in the series of Eq.~(\ref{eq:V2KAM}) or Eq.~(\ref{eq:V2KAMad}) for $\ep^2 V_2(t)$.

We now proceed to the construction of the second KAM iteration as described above.
The unitary transformation $T_2(t)$ is such that
\BEA
\label{eq:UH2UH3KAM}
T_2^\dagger(t)U_{H_2}(t,t_0)T_2(t_0)
= U_{H_3}(t,t_0),
\EEA
where  the propagator $U_{H_3}(t,t_0)$ is  associated with the Hamiltonian 
\BEA
H_3(t) \!\!&\equiv&\!\! H\efd(t) + \ep^4 V_3(t) ,\qquad
\label{eq:H3KAM}
\EEA
and can be expressed as
\BEA
\label{eq:UH3P3KAM}
U_{H_3}(t,t_0) = U_{H\efd}(t,t_2^{\prime\prime}) U_{R_3}(t,t_{0};t_2^{\prime\prime})U_{H\efd}(t_2^{\prime\prime},t_0) . \qquad
\EEA
The new effective Hamiltonian can be decomposed as
\BE
H\efd(t)\equiv H\efu(t) +\ep^2 D_2(t),\qquad
\EE
and its propagator accordingly written in the form
\BEA
U_{H\efd}(t,t_0)=U_{H\efu}(t,t_2)U_{P_2}(t,t_0;t_2)U_{H\efu}(t_2,t_0).
\EEA
From Eq.~(\ref{eq:interaction}) one deduces that  $U_{P_2}(t,t_0;t_2)$ is associated with the Hamiltonian
\BE
\ep^2 P_2(t;t_2)\equiv  U_{H\efu}(t_2,t)D_2(t)U_{H\efu}(t,t_2) ,
\EE
where $U_{H\efu}(t,t_0)$ is given  by Eq.~(\ref{eq:UHe1D1KAM}).
The corresponding Schr\"odinger equation  is straightforwardly integrated if 
 $D_2(t)$ is taken as
\BEA
\label{eq:D2KAM}
D_2(t)\!&\!\equiv \!&\!U_{H\efu}(t,t_2;t_1)D_2(t_2;t_2)U_{H\efu}(t_2,t;t_1) \NN\\
\!&\! \equiv \!&\! D_2(t;t_2) ,\quad
\EEA
with $D_2(t_2;t_2)$ arbitrary.
Two appealing cases are
\BES
\BEA
D_2(t_2;t_2)\!\!&\equiv &\!\!0,\quad \\
\label{eq:D2V2KAM}
D_2(t_2;t_2)\!\!&\equiv \!\!&  V_2(t_2).
\EEA
\EES
One then obtains
\BEA
\label{eq:S2KAM}
U_{P_2}(t,t_0;t_2)= e^{-i (t-t_0) \ep^2 D_2(t_2;t_2)}.
\EEA
The new effective propagator can be rewritten using Eq.~(\ref{eq:D2KAM}) as
\BEA
U_{H\efd}(t,t_0)\!&\!\equiv \!&\!
U_{H_0}(t,t_0)  e^{-i (t-t_0) \ep D_{1}(t_0;t_1)}e^{-i (t-t_0) \ep^2 D_{2}(t_0;t_2)} \NN\\
\!&\!\equiv \!&\! U_{H\efd}(t,t_0;t_1,t_2).
\label{eq:UHe2D2KAM}
\EEA

We now come to the definition of $T_2(t)$ given in Eq.~(\ref{eq:UH2UH3KAM}) and require that $\ep^4 V_3(t)$ contain no term of order lower than $\ep^4$.
Inserting $T_2(t)$ into the Schr\"odinger equation for $U_{H_3}(t,t_0)$ one arrives at
\BEA
\label{eq:coho2KAM}
\ep^4 V_3(t)=T_2^\dagger(t)H_2(t)T_2(t)- H\efd(t) -T_2^\dagger(t) i\frac{\partial}{\partial t}T_2(t). \quad
\EEA
We write  
\BEA
\label{eq:T2KAM}
T_2(t)\equiv e^{-i\ep^2 W_2(t)} ,
\EEA
where $W_2(t)$ is allowed to depend on $\ep$ (although we do not indicate it explicitly).
Substituting into Eq.~(\ref{eq:coho2KAM}) and
requiring that all terms of order lower than $\ep^4$ vanish identically leads to the  differential equation 
\BEA
\label{eq:W2dotKAM}
\frac{\partial}{\partial t}  W_2(t) =V_2(t)-D_2(t)+i\left[W_2(t),H\efu(t)\right] .
\EEA
Its general solution reads
\BEA
\label{eq:W2KAM}
W_2(t)\!&\!=\!&\!\int_{t_2^{\prime}}^t d u U_{H\efu}(t,u)\left[V_2(u)-D_2(u;t_2)\right]U_{H\efu}(u,t)\NN\\
\!&\!+\!&\!U_{H\efu}(t,t_0)B_2 U_{H\efu}(t_0,t) , \quad
\EEA
where  $B_2$ is any constant self-adjoint operator.
Here we shall set $B_2$ to 0.

The propagator $U_{H_1}(t,t_0)$ we are looking for is obtained from Eqs.~(\ref{eq:UH1UH2KAM}) and (\ref{eq:UH2UH3KAM})
\BEA
\label{eq:UH1UH3KAM}
U_{H_1}(t,t_0)={T_1}(t){T_2}(t)
U_{H_3}(t,t_0)T_2^{\dagger}(t_0)T_1^{\dagger}(t_0) .\qquad
\EEA
Substituting Eq.~(\ref{eq:UH3P3KAM}) and neglecting $U_{R_3}(t,t_0;t_2^{\prime\prime})$ since it is associated with a Hamiltonian of order $\ep^4$, we find
\BEA
U_{H_1}(t,t_0)=U_{H_1}^{(2)}(t,t_0)+\mathcal{O}(\ep^4) , \qquad
\label{eq:UH1UH12KAM}
\EEA
with the two-iteration KAM expansion
\BEA
 U_{H_1}^{(2)}(t,t_0) = {T_1}(t){T_2}(t)U_{H\efd}(t,t_0)
 T_2^\dagger(t_0)T_1^\dagger(t_0).\label{eq:UH12KAM}
\EEA
This expansion features two additional arbitrary times ($t_2$ and $t_2^\prime$) and two additional arbitrary operators ($D_2(t_2;t_2)$ and $B_2$) that can be chosen so as to improve its accuracy.
The structure of the equations and of the operators involved is exactly the same for the second KAM  iteration as for the first one, and will be the same for any iteration.
In particular,  the effective perturbation determined at each step is always of the form of Eq.~(\ref{eq:V2KAMad}) since each step is just a renaming of the previous one.
This is in contrast to the order by order methods where the determination of that operator requires each time more algebra.

\subsubsection{Summary of the time-dependent KAM algorithm}
\label{sec:summary}
The propagator of the perturbed Hamiltonian $H_1(t)=H_0(t)+\ep V_1(t)$ is approximated after $n$ iterations according to
\BE
U_{H_1}(t,t_0)=U_{H_1}^{(n)}(t,t_0)+\mathcal{O}(\epn),
\EE
with the unitary $n$-iteration KAM expansion 
\BEA
U_{H_1}^{(n)}(t,t_0)\!&\! =   \!&\!{T_1}(t)\ldots T_{n}(t)U_{H\efn}(t,t_0) \NN\\\!&\! \times \!&\! T_{n}^\dagger(t_0)\ldots T_1^\dagger(t_0).
\label{eq:UH1nKAM}
\EEA
We recall that the superexponential character of the correction terms stems from the fact that each KAM transformation reduces the perturbation to a new effective perturbation whose order is squared.
One defines
\BEA
U_{H\efn}(t,t_0)\!&\!\equiv \!&\!
U_{H_0}(t,t_0)  \exp \left[-i (t-t_0) \ep D_{1}(t_0;t_1)\right]\NN\\
 \times \cdots \!&\!  \times \!&\!  \exp \left[-i (t-t_0) \epnm D_{n}(t_0;t_n)\right], \qquad
\label{eq:UHenKAM}
\EEA
with  the possibility to choose one of the following for any $k\geq 1$ 
\BES
\label{eq:DnsKAM}
\BEA
\label{eq:Dn0KAM}
D_{k}(t;t_k)\!\!&\equiv \!\!& 0, \\
\label{eq:DnVnKAM}
D_{k}(t;t_k)\!\!& \equiv \!\!& U_{H\efkm}(t,t_{k}) V_{k}(t_{k}) U_{H\efkm}(t_k,t). \qquad
\EEA
\EES
Each KAM transformation reads 
\BEA
\label{eq:TnKAM}
T_{k}(t)\equiv \exp \left[-i\epkm W_{k}(t)\right] , \qquad
\EEA
where 
\BE
W_{k}(t)\equiv \int_{t_{k}^{\prime}}^t \! d u U_{H\efkm}(t,u)\left[V_{k}(u)-D_{k}(u;t_k)\right]U_{H\efkm}(u,t).
\label{eq:WnKAM}
\EE
One has $U_{H\efz}(t,t_0)\equiv U_{H_0}(t,t_0)$.
To continue the algorithm we determine the new perturbation 
$\epn V_{n+1}(t)$ in terms of the operators $W_n(t)$, $V_n(t)$ and $D_n(t;t_n)$ obtained at the preceding step
\BE
\label{eq:VnKAM}
V_{n+1}\equiv \sum_{k=1}^{\infty}
\frac{i^{k}(\epnm)^{k-1}}{(k+1)!}\,\ad^{k}\left(W_n,kV_n+D_n\right),
\EE
with the usual definition of $\ad(A,B)$ recalled in Eq.~(\ref{eq:ad}) and where the infinite series can be truncated to a prescribed order.
This new effective perturbation  has exactly the same structure at each iteration which is useful for applications, particularly when high orders computations are needed.

At each KAM iteration two arbitrary times $t_k$ and $t_k^\prime$ are introduced through Eqs.~(\ref{eq:DnsKAM}) and (\ref{eq:WnKAM}). 
As a consequence, one has the following dependence on these free parameters:
\BEA
U_{H\efn}(t,t_0) \!&\!\equiv \!&\! U_{H\efn}(t,t_0;t_1,\cdots,t_n),\NN\\
T_{n}(t)  \!&\!\equiv \!&\! T_{n}(t;t_1,\cdots,t_n,t_1^\prime,\cdots,t_n^\prime),\NN\\
W_{n}(t) \!&\!\equiv \!&\!
 W_{n}(t;t_1,\cdots,t_n,t_1^\prime,\cdots,t_n^\prime),\NN\\
 V_{n}(t) \!&\!\equiv \!&\!
 V_{n}(t;t_1,\cdots,t_{n-1},t_1^\prime,\cdots,t_{n-1}^\prime).
 \label{eq:dependence}
\EEA
These quantities, together with the choice of Eq. (\ref{eq:Dn0KAM})
or (\ref{eq:DnVnKAM}) for the arbitrary operator $D_n(t_n;t_n)$, may significantly affect the accuracy of the $n$-iteration KAM expansion. 

\subsubsection{Comparison with the Magnus expansion}
\label{sec:comp}
The Magnus and KAM expansions differ in several respects.
First it is remarkable that the KAM algorithm can be implemented in the original representation. In Appendix~\ref{app:KAM} we show that the result obtained for the KAM expansion in the interaction representation is identical at any level of approximation.

Most importantly of course is the superexponential character of the KAM expansion which manifests itself as of the second iteration.
However, the first iteration of these algorithms are generally different, owing to the non-commutativity of the operators involved.
Indeed, Eq.~(\ref{eq:UH1nMag}) gives for the first order Magnus expansion
\BE
\label{eq:UH1MagM1}
U_{H_1}^{(1)}(t,t_0)=e^{-i\ep M_1(t;t_0)} U_{H_0}(t,t_0),
\EE
while the one-iteration KAM expansion of Eq.~(\ref{eq:UH11KAM})  reads
\BEA
\label{eq:UH1KAMM1}
U_{H_1}^{(1)}(t,t_0)\!&\!=\!&\!e^{-i\ep \left\{ M_1(t;t_0)- M_1(t_1^{\prime};t_0)+(t_1^{\prime}-t)D_1(t;t_1)\right\}}
\NN\ \\ \!&\!\times \!&\!  e^{-i\ep(t-t_0)D_1(t;t_1)}e^{-i\ep \left\{  M_1(t_1^{\prime};t_0)+(t_0-t_1^{\prime})D_1(t;t_1)\right\}}\NN\\ \!&\! \times \!&\!  U_{H_0}(t,t_0).
\EEA
We recall from Eq.~(\ref{eq:MnMag}) that 
\BEA
\label{eq:M1Mag}
 M_1(t;t_0)=\int_{t_0}^t \!d u  U_{H_0}(t,u) V_1(u)U_{H_0}(u,t).
\EEA
To compare these expressions we shall cast the product of the exponentials of Eq.~(\ref{eq:UH1KAMM1})  into a single exponential using the Campbell-Baker-Hausdorff formula \cite{Hausdorff,Pechukas}
\BES
\BE
e^{A}e^{B}=e^C,
\EE
where
\BE
\label{eq:BH}
C=A+B+\frac{1}{2}\left[A,B\right]+\frac{1}{12}\left[A-B,\left[A,B\right]\right]+\cdots .
\EE
\EES
We note that the exponents of Eq.~(\ref{eq:UH1KAMM1}) precisely sum up  to that of Eq.~(\ref{eq:UH1MagM1}), i.e., $\ep M_1(t)$.
Hence, by Eq.~(\ref{eq:BH}), these expansions differ by terms of order $\ep^2$.
In other words, these expansions differ through terms whose order is that of their remainder, which therefore enables one to recover precisely the same expansion up to a given order.
To compute explicitly these terms of order $\ep^2$ and show that they generally do not  vanish we apply the Campbell-Baker-Hausdorff formula twice to reduce the three exponentials of  Eq.~(\ref{eq:UH1KAMM1}) to a single one
\BEA
\label{eq:UH1KAMM1BH}
U_{H_1}^{(1)}(t,t_0)\!&\!=\!&\!e^{-i\ep M_1(t;t_0)-\frac{\ep^2}{2}  K_1(t;t_0,t_1,t_1^\prime)+O(\ep^3) }\NN\\
\!&\! \times \!&\! U_{H_0}(t,t_0),
\EEA
where 
\BEA
 K_1(t;t_0,t_1,t_1^\prime)\!&\! \equiv \!&\!  (t-t_1^{\prime})\left[M_1(t;t_0),D_1(t;t_1)\right] \quad\NN\\
\!&\!+\!&\!(t_0-t)\left[M_1(t_1^{\prime};t_0),D_1(t;t_1)\right]\NN\\ \!&\!+\!&\!\left[M_1(t;t_0),M_1(t_1^{\prime};t_0)\right].\qquad
\EEA
We recall that, according to Eq.~(\ref{eq:D1sKAM}), we choose either  $D_1(t;t_1)=0$ or $D_1(t;t_1)=U_{H_0}(t,t_1) V_1(t_1) U_{H_0}(t_1,t)$.
As a consequence $K_1(t;t_0,t_1,t_1^\prime)$ is generally non zero and therefore, at this first level of approximation, the KAM and Magnus techniques differ by terms of order $\ep^2$.
However, if one chooses $D_1(t;t_1)=0$ together with $t_1^\prime=t_0$, then the one-iteration KAM expansion and the first order Magnus expansion coincide (this will no longer be true for the next levels of approximation). 

Note that for the KAM algorithm, as  we discuss in the following section, the choice of $D_1(t;t_1)$ with an arbitrary $t_1$ allows one to enhance the convergence precisely by acting on these higher order terms (we emphasize that these terms appear here as higher order ones because of the use of the Campbell-Baker-Hausdorff formula; but at the level of Eq.~(\ref{eq:UH1KAMM1}), each exponent is indeed of order $\ep$).

For higher orders and iterations however the KAM algorithm is {\em a priori} expected to perform far better owing to both the superexponential character and the possibility to enhance the accuracy.

\section{Improving the accuracy}
\label{sec:accuracy}
The formulation of the time-dependent KAM technique presented in Sec.~\ref{sec:KAM}  reveals the existence of several degrees of freedom which are at our disposal to  reduce the error
without prior knowledge of the exact solution: i) the choice of Eq.~(\ref{eq:DnsKAM}) for the operators $D_k(t;t_k)$, and ii)  the free parameters $t_k$ and $t_k^{\prime}$ of Eqs.~(\ref{eq:DnVnKAM}) and (\ref{eq:WnKAM}).
There is also a third way discussed below: iii) the possibility to consider another identification of the perturbation and unperturbed Hamiltonian.

The items i) and ii) also apply to the Poincar\'e-Von Zeipel and the Van Vleck techniques, although to a smaller extent, as we describe below. 

\subsection{Choice of $D_k(t;t_k)$ and correspondence between resonances and secular terms.}
\label{sec:limit}
Each iteration of the time-dependent Poincar\'e-Von Zeipel, the time-dependent Van Vleck and the time-dependent KAM algorithms features an arbitrary operator $D_k(t_k;t_k)$.
In the preceding section, in addition to the simplest case $D_k(t_k;t_k)=0$, we suggested the choice $D_k(t_k;t_k)=V_k(t_k)$ where $t_k$ is an arbitrary time.

The first iteration involves the operator $D_1(t;t_1)$ which satisfies the same equation in the three algorithms, namely Eq.~(\ref{eq:D1KAM}).
This equation is actually the general solution to the differential equation
\BE
\label{eq:D1KAMdiff}
[H_0(t),D_1(t;t_1)]=i \frac{\partial}{\partial t} D_1(t;t_1).
\EE
This latter equation, together with Eq.~(\ref{eq:W1dotKAM}) for $W_1(t)$, are the time-dependent generalization of the so-called cohomology equations \cite{PhysicaA} considered in the stationary case. 

The \textit{time-independent problem}, i.e. the problem of finding a
transformation $T_{1}$ that enables one to simplify the time-independent
Hamiltonian $H_{1}$ according to $T_{1}^{\dagger }H_1T_{1}=H_{0}+\ep D_{1}+\ep%
^{2}V_{2}$, is recovered when one conveniently chooses  $T_{1}$ as time-independent. 
This transformation is sometimes called contact transformation \cite{Shaffer} or level-shift transformation \cite{Primas}.

In this case all the operators, and in
particular $D_{1}$ and $W_{1}$, are
time-independent and the standard cohomology equations are recovered
\BES
\BEA
&&\left[
H_{0},D_{1}\right] =0,\\
&&V_{1}-D_{1}+i\left[ W_{1},H_{0}\right] =0. 
\EEA
\EES
Their solutions can be determined using the
following key property \cite{PhysicaA}: $W_{1}$ exists if and only if $\Pi
_{H_{0}}(D_{1}-V_{1})=0$, where $\Pi _{H_{0}}$ is the projector in the
kernel of the application $A\mapsto \lbrack A,H_{0}]$ (for an operator $A$
acting on the same Hilbert space as $H_{0}$). The projector $\Pi _{H_{0}}$
applied on an operator $A$ captures thus all the part $B$ of $A$ which
commutes with $H_{0}$: $[B,H_{0}]=0$. 
The unique solution $D_{1}$ allowing $%
W_{1}$ to exist and satisfying Eq. (\ref{eq:D1KAMdiff}) is thus%
\begin{equation}
D_{1}=\Pi _{H_{0}}V_{1}\equiv \lim_{T\rightarrow \infty }\frac{1}{T}%
\int_{0}^{T}e^{-itH_{0}}V_{1}e^{itH_{0}}.  \label{D1}
\end{equation}%
The \textit{resonances} are associated with terms of $V_{1}$ which commute
with $H_{0}$. 
Application of Eq. (\ref{D1}) can be interpreted as an \textit{%
averaging }of $V_{1}$ with respect to $H_{0}$ which allows one to extract resonances.

For the \textit{time-dependent problem}, the general solution to Eq. (\ref%
{eq:W1dotKAM}) is given by Eq.~(\ref{eq:W1KAM}).
Defining the average 
\begin{equation}
\Pi _{-}V_{1}\equiv \lim_{\tau \rightarrow \infty }\frac{1}{\tau }%
\int_{t-\tau }^{t}dsU_{H_{0}}(t,s)V_{1}(s)U_{H_{0}}(s,t),  \label{PiK0}
\end{equation}%
one can show the following property: if $W_1(t)$ is bounded
for negative infinite times, then $\Pi _{-}(V_{1}-D_{1})=0$. This is
satisfied by $D_{1}=\Pi _{-}V_{1}$, the only solution compatible with Eq. (\ref{eq:D1KAMdiff}) and the projector $\Pi _{-}$. 
This means that the averaging $D_{1}=\Pi _{-}V_{1}$ allows to 
remove secular terms at negative infinite times. We remark that
this definition of the average, Eq.~ (\ref{PiK0}), can in fact be recovered from
the formal calculation of the average $\Pi _{K_{0}}V_{1}$ of Eq.~(\ref{D1})
with respect to $K_{0}=-i\frac{\partial }{\partial t}+H_{0}$ in an extended space, which includes time as a coordinate \cite{SchererIII,art1}. 
This gives the precise correspondence between the resonances of stationary problems and the secular terms of time-independent problems.

In Ref.~\cite{art1} it was shown for perturbations that are localized in time, in a finite but possibly large interval $t_{i}\leq t \leq t_{f}$, that Eq.~(\ref{PiK0}) reduces to 
\BE%
\Pi _{-}V_{1}=U_{H_{0}}(t,t_{i})V_1(t_{i})U_{H_{0}}(t_{i},t).
\EE
This is a
particular solution to Eq. (\ref{eq:D1KAMdiff}) corresponding to the choice $t_1\equiv t_i$ in Eqs.~(\ref{eq:D1KAM}) and (\ref{eq:D1V1KAM}).
An alternate definition of the average
\BE
\Pi _{+}V_{1}\equiv \lim_{\tau
\rightarrow \infty }\frac{1}{\tau }\int_{t}^{t+\tau
}dsU_{H_{0}}(t,s)V_{1}(s)U_{H_{0}}(s,t),
\EE  gives a different result
\BE
\Pi
_{+}V_{1}=U_{H_{0}}(t,t_{f})V_1(t_{f})U_{H_{0}}(t_{f},t),
\EE
 and allows one to 
remove secular terms at positive infinite times. Generally one cannot remove simultaneously the secular terms at negative and positive large times. 
This shows a conceptual difference between stationary
resonances and secular terms associated with perturbations localized in time. 
Furthermore, it suggests that combining both definitions in a non-trivial way 
gives a new secular term that could improve the convergence of the algorithm.
This is achieved by the general solution  to Eq. (\ref{eq:D1KAMdiff}), Eq.~(\ref{eq:D1KAM}), written with the perturbation evaluated at a free time $t_{1}$ as
the arbitrary operator [cf. Eq.~(\ref{eq:D1V1KAM})]
\begin{equation}
D_{1}(t;t_1)\equiv U_{H_{0}}(t,t_{1})V_{1}(t_{1})U_{H_{0}}(t_{1},t).  \label{D1opt}
\end{equation}%
The free $t_{1}$ can then be chosen as we describe below to minimize the remainder obtained at  the first iteration of the perturbative algorithm.

The next iterations also  offer the possibility to choose for  $D_k(t;t_k)$, in particular $0$ or $U_{H\efkm}(t,t_{k}) V_{k}(t_{k}) U_{H\efkm}(t_k,t)$ similarly to Eq.~(\ref{D1opt}). 
Note however that for the Poincar\'e-Von Zeipel  and the Van Vleck techniques, the choice corresponding to $V_k(t_k)$ can only be made once (for an arbitrary value of $k$).
As explained in Appendixes A and B, respectively, this is due to the order by order character of these techniques.

\subsection{Enhancing the convergence}
\label{sec:optim}
After one iteration of the KAM algorithm one deduces from  Eqs.~(\ref{eq:UH1UH2KAM}) and (\ref{eq:UH2KAM}) an  exact expression for the full propagator
\BEA
\label{eq:UH1P2KAMbis}
U_{H_1}(t,t_0)\!&\!=\!&\!{T_1}(t;t_1,t_1^\prime)U_{H\efu}(t,t_1^{\prime\prime};t_1)U_{R_2}(t,t_0;t_1,t_1^\prime,t_1^{\prime\prime})\NN\\
\!&\! \times \!&\! U_{H\efu}(t_1^{\prime\prime},t_0;t_1)T_1^{\dagger}(t_0;t_1,t_1^\prime),\qquad 
\EEA
where $t_1$, $t_1^\prime$ and  $t_1^{\prime\prime}$ are arbitrary times [cf. Eq.~(\ref{eq:dep1})].
The propagator $U_{R_2}(t,t_0;t_1,t_1^\prime,t_1^{\prime\prime})$ of the Hamiltonian given in Eq.~(\ref{eq:R2KAM}) is associated with a second order generator
\BE
U_{R_2}(t,t_0;t_1,t_1^\prime,t_1^{\prime\prime})\equiv e^ {-i\ep^2 G_2(t;t_0,t_1,t_1^\prime,t_1^{\prime\prime})}.\quad  \quad
\EE 
To obtain the one-iteration KAM expansion $U_{H_1}^{(1)}(t,t_0)$ we neglected this propagator, replacing it by the identity  in the above product.
Obviously, the closer $U_{R_2}(t,t_0;t_1,t_1^\prime,t_1^{\prime\prime})$ is to the identity, the smaller the correction terms are, i.e., the more accurate the approximation is.
We can improve this accuracy if we can  make
 that propagator closer to the identity, or equivalently, its generator closer to zero.
The distance is defined through the norm  $||A||=\max_{||\psi||=1} ||A \psi||$ with $\psi$ in the Hilbert space of the problem.
For an Hermitian matrix, this norm is the largest absolute value of its eigenvalues.

We calculate this generator $\ep^2 G_2(t;t_0,t_1,t_1^\prime,t_1^{\prime\prime})$ by solving the Schr\"odinger equation  with the Hamiltonian of Eq.~(\ref{eq:R2KAM})  in the form of an exponential using Eq.~(\ref{eq:UH1KAMM1BH}). 
This is a time-dependent problem with a zero unperturbed Hamiltonian and whose  perturbation is $\ep^2 U_{H\efu}(t_1^{\prime\prime},t;t_1) V_2(t;t_1,t_1^\prime) U_{H\efu}(t,t_1^{\prime\prime};t_1)$.
Hence, we evaluate the lowest order contribution to  $G_2(t;t_0,t_1,t_1^\prime,t_1^{\prime\prime})$ as
\BEA
\label{eq:G2}
G_2^{(2)}(t;t_0,t_1,t_1^\prime)\equiv \int_{t_0}^t du \!&\!\!&\! U_{H\efu}(t_0,u;t_1)  V_2(u;t_1,t_1^\prime)\NN\\
\times  \!&\!  \!&\!  U_{H\efu}(u,t_0;t_1)  ,\qquad
\EEA
where we set $t_1^{\prime\prime}=t_0$ (its precise value is not relevant since the one-iteration KAM expansion $U_{H_1}^{(1)}(t,t_0;t_1,t_1^\prime)$ is independent of the parameter $t_1^{\prime\prime}$).
It is this operator $G_2^{(2)}(t;t_0,t_1,t_1^\prime)$ that has to remain small for the algorithm to converge \footnote{In a similar vein, on the basis of Eq.~(\ref{eq:UH1KAMM1BH}) one can define a small parameter $\ep m_1$ where $\ep$ is the perturbative (ordering) parameter [cf. Eq.~(\ref{eq:M0M1})] and $m_1$ is the norm of  $M_1(t,t_0)$. Notice from the definition given in Eq.~(\ref{eq:M1Mag}) that this operator samples the perturbation $V_1(t)$ during its whole duration.}.
Having the arbitrary times $t_1$ and $t_1^\prime$ at our disposal, we can actually enhance the convergence of the algorithm by minimizing the norm of   this operator with respect to these free parameters.

Similarly, the $n$-iteration KAM expansion of Eq.~(\ref{eq:UH1nKAM}) can be optimized by minimizing the norm of the following operator with respect to one or several of the free parameters $t_1,\cdots,t_{n},t_1^\prime,\cdots,t_{n}^\prime$
\BEA
\label{eq:Gn}
G_{n+1}^{(n+1)}(t) \!&\! \equiv \!&\! \int_{t_0}^t du U_{H\efn}(t_0,u) V_{n+1}(u) U_{H\efn}(u,t_0)\NN\\
\!&\! \equiv \!&\!  G_{n+1}^{(n+1)}(t;t_1,\cdots,t_{n},t_1^\prime,\cdots,t_{n}^\prime)  .\qquad
\EEA
The dependence of $U_{H\efn}(t,t_0)$ and $V_{n+1}(t)$ on these parameters is given in Eq.~(\ref{eq:dependence}).

It turns out, as will be illustrated in Sec.~\ref{sec:illu}, that modifying the parameters $t_k$ and/or $t_k^{\prime}$ can improve the accuracy by more than one order of magnitude already for $k=1$.

\subsection{KAM expansion with another identification of the unperturbed Hamiltonian}
\label{sec:newid}
 The perturbed Hamiltonian can be written as
\BEA
 H_1(t) \!\!&=&\!\! \underbrace{H_0(t)+\ep D_1(t;t_1)}_{}+ \underbrace{\ep V_1(t) -\ep D_1(t;t_1)}_{}\NN\\
 \!\!&\equiv&\!\!\qquad H_0^{\prime}(t;t_1)\qquad +\qquad\ep V_1^{\prime}(t;t_1)  ,\quad\qquad 
 \label{eq:H1prime}
 \EEA
where
$D_1(t;t_1)\equiv U_{H_0}(t,t_1)V_1(t_1)U_{H_0}(t_1,t)$ with $t_1$ arbitrary [cf. Eqs.~(\ref{eq:D1KAM}) and (\ref{eq:D1V1KAM})].
The propagator associated with $H_0^{\prime}(t;t_1)$ can always be determined since  by Eqs.~(\ref{eq:He1KAM}) and (\ref{eq:UHe1D1KAM}) one has
\BES
\BEA
 \label{eq:H0prime}
H_0^{\prime}(t;t_1) &=& H\efu(t;t_1), \qquad \\
U_{H_0^{\prime}}(t,t_0;t_1)&= & U_{H\efu}(t,t_0;t_1).\qquad
 \label{eq:UH0prime}
\EEA
\EES

We now apply the KAM algorithm exactly as summarized in Sec.~\ref{sec:summary}, i.e. with the same definitions for all the operators involved in the expansion, but with the identification of Eq.~(\ref{eq:H1prime}).
This decomposition has the property that $V_1^{\prime}(t_1;t_1)=0$ which implies by Eq.~(\ref{eq:D1V1KAM}) that $D_1^{\prime}(t;t_1)=0$. 
The free parameter $t_1$  is therefore introduced here  through Eq.~(\ref{eq:UH0prime}).
One arrives at a KAM expansion which is still of the form of Eq.~(\ref{eq:UH1nKAM}) but may significantly differ from that resulting from the conventional decomposition.

We emphasize that the possibility to consider the  identification of Eq.~(\ref{eq:H1prime}) as a new starting point for a perturbative treatment is specific to the KAM technique which is not an order by order method, contrary to the Magnus, the Poincar\'e-Von Zeipel and the Van Vleck algorithms.
\section{Beyond the sudden approximation}

\label{sec:sudden}
\subsection{Preliminaries}
\label{sec:preli}
We consider a system described by the Hamiltonian ${\sf H}$ (autonomous or not).
It is perturbed by a time-dependent Hamiltonian ${\sf V(t)}$ whose characteristic duration is $\tau$.
This latter quantity is the time during which the interaction differs significantly from zero, and not the full duration of the interaction, which may be large but finite. 
Here we define $\tau$ as twice the full width at half maximum, having in mind a perturbation which presents a time-dependent envelope.
We assume that the perturbation ${\sf V(t)}$ satisfies 
\BE
\label{eq:hypV}
[{\sf V(t),V(t_0)}]=0 \quad \forall {\sf t,t_0},
\EE
which is realized in many situations of physical interest.
The propagator of the perturbed system evolves according to the Schr\"odinger equation
\begin{equation}
\label{eq:Us}
i\hbar {\sf \ddpt U(t,t_{0}) = \left\{ H+  V(t) \right\}
U(t,t_{0})},
\end{equation}
with ${\sf U(t_0,t_{0})}=\un$, the identity operator on the appropriate Hilbert space $\Hs$.
We define a dimensionless time $t$ and dimensionless operators $H$, $V(t)$ and $U(t,t_0)$ through
\BEA
{\sf t}&\equiv &\tau t,\NN\\
{\sf H}&\equiv &\hbar \omega H,\NN\\
{\sf V(t)}&\equiv &\frac{\hbar}{\tau} V(t),\NN\\
{\sf U(t,t_{0})}&\equiv &U(t,t_0),
\EEA
where $\omega$ is some characteristic frequency of ${\sf H}$.
In dimensionless units Eq.~(\ref{eq:Us}) becomes
\begin{equation}
\label{eq:Usad}
i\ddpt U(t,t_{0}) = \left\{ V(t)+ \ep H \right\}
U(t,t_{0}),
\end{equation}
where we define a {\em sudden parameter} $\ep \equiv \omega \tau$.
 
The sudden or impulsive regime corresponds to the limit $\ep \rightarrow 0$.
Our aim is to obtain a perturbative expansion for  the evolution operator
$U(t,{t}_{0})$ of the perturbed Hamiltonian
$V(t)+\ep H$  beyond the sudden regime.
To this aim we identify the original perturbation $V(t)$ as the unperturbed Hamiltonian $H_0(t)$ and the original unperturbed Hamiltonian $H$ as the perturbation $V_1$
\BES
\label{eq:H0V1soudain}
\BEA
H_0(t)&\equiv & V(t),\\
V_1 &\equiv & H. \label{eq:V1H}
\EEA
\EES 
Note that we need only consider a finite interval of time as $V(t)$ is localized in time.
By virtue of Eq.~(\ref{eq:hypV}), the propagator for $\ep=0$ reads
\begin{equation}
\label{eq:UH0V}
U_{H_0}(t,t_{0}) = \exp\left\{-i \int_{t_{0}}^{t} V(u) \, d u\right\}.
\end{equation}
We shall follow this approach below and  consider  the various perturbative schemes described in Sec.~\ref{sec:time} in the case of two-level systems driven by short pulses. 

\subsection{Illustration on pulse-driven two-level systems}
\label{sec:illu}
Our purpose is to compare the various algorithms and  investigate the convergence enhancement as well as to show that the unitary  time-dependent KAM theory is well suited to study regimes beyond the impulsive or sudden limit.

In the notations of the preliminaries, we take
${\sf H}=\hbar \omega \sigma_3$ and ${\sf V(t)}=\frac{\hbar}{\tau} \Omega(t)\sigma_1$ where $\Omega(t)$ is a pulse shape function and
 $\sigma_k$  are the  Pauli matrices
\BE
\sigma_1=\begin{pmatrix}
0 & 1 \\ 1 & 0
\end{pmatrix},\quad \sigma_2=\begin{pmatrix}
0 & -i \\ i & 0
\end{pmatrix}, \quad \sigma_3=\begin{pmatrix}
1 & 0 \\ 0 & -1
\end{pmatrix}.\NN
\EE 
Hence, the operators defined in Eq.~(\ref{eq:H0V1soudain}) are here
$H_0(t)=\Omega(t) \sigma_1$ and $V_1=\sigma_3$.
The Schr\"odinger equation reads
\begin{equation}
\label{eq:Us2}
i\ddpt U_{H_1}(t,t_{0}) = \left[\Omega(t)\sigma_1+\epsilon \sigma_3 \right]
U_{H_1}(t,t_{0}),   \quad
\end{equation}
with $U_{H_1}(t_0,t_{0}) =\un_{\Cs}$.
For  $\epsilon=0$ its solution  is
\BE
\label{eq:U0}
U_{H_0}(t,t_{0}) \equiv e^{-iA(t)\sigma_1} ,
\EE
where $A(t)\equiv\int_{t_0}^t \Omega(u) \, d u$.
The pulse area $A\equiv A(\infty)$ is a dimensionless parameter that can be fixed independently of the sudden parameter $\ep$ $(\equiv \omega \tau)$ that we take here as the perturbative parameter.
This allows us, in particular,  to treat large nonperturbative areas for short pulse durations.

The Magnus expansion $U_{H_1}^{(n)}(t,t_0)$ is given by Eqs.~(\ref{eq:UH1nMag})-(\ref{eq:VnMag}).

The KAM expansion is obtained from Eqs.~(\ref{eq:UH1nKAM})-(\ref{eq:VnKAM}).
We distinguish three types of KAM expansions, reflecting the choices discussed in Secs.~\ref{sec:KAM}, \ref{sec:limit} and \ref{sec:newid}:
\begin{itemize}
\item[Type] A: each iteration  $k$ involves the operator $D_k(t_k;t_k)=0$ [cf. Eq.~(\ref{eq:Dn0KAM})].
\item[Type]  B: $D_k(t_k;t_k)=V_k(t_k)$ for all $k$  [cf. Eq.~(\ref{eq:DnVnKAM})].
\item[Type]  C: the unperturbed Hamiltonian is defined as $H_0^{\prime}(t;t_1)=H_0(t)+\ep U_{H_0}(t,t_1) V_1(t_1)U_{H_0}(t_1,t)$ [cf. Eq.~(\ref{eq:H1prime})].
\end{itemize}
For the type A $n$-iteration expansion one has $n$ free times
\BE
U_{H_1}^{(n)}(t,t_0)=U_{H_1}^{(n)}(t,t_0;t_1^\prime,\cdots, t_n^\prime),\NN
\EE
while for the types B and C one has $2 n$ such parameters
\BE
U_{H_1}^{(n)}(t,t_0)=U_{H_1}^{(n)}(t,t_0;t_1,\cdots, t_n,t_1^\prime,\cdots, t_n^\prime).\NN
\EE

In the case of two-level systems, the infinite series of Eq.~(\ref{eq:VnKAM}) for the new effective KAM perturbation $V_{n+1}(t)$ can be cast into the form \cite{art1}

\BEA
V_{n+1}\!&\!=\!&\!\left[W_n,a_nV_n+b_nD_n\right]\NN\\
\!&\!+\!&\!
\ep_n \left[W_n,\left[W_n,c_nV_n+ d_nD_n\right] \right], \qquad \label{eq:VnKAMresom}
\EEA
where 
\BEA
a_n\!\!&\equiv &\!\!i\frac{\cos{\ep_n \lambda_n}+\ep_n \lambda_n \sin{\ep_n \lambda_n}-1}{\ep_n^2 \lambda_n^2},\quad
b_n\equiv i\frac{1-\cos{\ep_n \lambda_n}}{\ep_n^2 \lambda_n^2},\NN\\
c_n\!\!&\equiv &\!\!\frac{\ep_n \lambda_n \cos{\ep_n \lambda_n}-\sin{\ep_n \lambda_n}}{\ep_n^3 \lambda_n^3},\quad
d_n\equiv c_n+i b_n, \label{eq:abcd}
\EEA
with $\ep_n\equiv\epnm$ and  $\lambda_n(t)\equiv \sqrt{-\det W_n(t)}$.

For given $\ep$ and $A$, the error $\Delta_n$ at the end of the pulse between the
 numerical solution of the Schr\"odinger
equation and the result obtained after $n$ iterations is defined as
\BEA
\label{Err}
\Delta_n\equiv ||U_{H_1}(t_f,t_i)-U^{(n)}_{H_1}(t_f,t_i)||. \quad
\EEA
We also define the error $\delta_n$ in the transition probability from the lower state $|-\rangle$ to the upper state $|+\rangle$
\BE
\delta_n  \equiv \left|\langle+| U^{(n)}_{H_1}(t_f,t_i)|-\rangle\right|^2
- \left|\langle+| U_{H_1}(t_f,t_i)|-\rangle\right|^2.
\EE

We consider the following dimensionless pulse shape between the dimensionless time $t_i=0$ and $t_f=1$
\BE
\label{pulse}
\Omega(t)=\left\{\begin{array}{cc}2 A\sin^2\left(\pi t\right)\quad &
 0\le t\le 1,\\
0 & \text{ elsewhere}.\end{array}\right.
\EE

Figure 1 displays the common logarithm of the error $\Delta_1$ and the error $\delta_1$ as a function of the pulse area $A$ for the Magnus expansion and the three types of one-iteration KAM expansions in the case $t_1=t_1^\prime=0$ and  $\ep=0.5$.
The errors $\Delta_1$ and $\delta_1$ globally decrease when $A$ increases.
This is expected on the basis of Eqs.~(\ref{eq:Us2}) and (\ref{pulse}) as the relative importance of the perturbation then decreases.
One also observes marked oscillations in $\Delta_1$ and $\delta_1$ with a pseudo-period of $\pi$.
This stems from the form of the unperturbed propagator [cf. Eq.~(\ref{eq:U0})] and the fact that it always appears twice, in particular in the operator 
$G_2^{(2)}(t;t_0)$ of Eq.~(\ref{eq:G2}) which controls the error after one iteration.

Let us recall that the one-iteration KAM expansion of type A coincides with the first order Magnus expansion for $t_1^\prime=0$.
It is seen, by both measures of the error, that each of the KAM expansions can perform better than the other ones on some intervals of $A$. 
Hence, in order to establish a fair comparison, we shall consider the particular value $A=1$ where the first order Magnus expansion and the one-iteration KAM expansion of type B yield essentially the same error $\Delta_1$ for $\ep=0.5$.
This remains true for all values of $\ep$ up to 2 as can be seen from Fig.~2 which depicts the errors $\Delta_1$ and $\delta_1$ as a function of $\ep$ for $A=1$.

In Fig.~2 we also present the one-iteration KAM expansion of  type B that is  optimized by choosing $t_1=0.5, t_1^\prime=0.22$.
We see that the error $\Delta_1$ is reduced (with  respect to the comparable Magnus and non-optimized type B KAM expansions) by more than one order of magnitude up to values of $\ep$ equal to unity.
The error $\delta_1$ on the transition probability  is also considerably reduced.
Notice that  the values of $\delta_1$ for the Magnus and non-optimized type B KAM expansions differ while the values of $\Delta_1$ are indistinguishable.
For comparison, we also consider the (non-unitary) Dyson expansion \cite{Pechukas}.
Recall that it is obtained by repeated use of the integral form of the Schr\"odinger equation in the interaction representation with respect to $H_0(t)$
\BEA
 U_{H\inte}(t,t_0;t_0)\!&\!=\!&\! \un-i\int_{t_0}^t du H\intu(u;t_0) U_{H\inte}(u,t_0;t_0)\NN\\
 \!&\!=\!&\! \un-i\int_{t_0}^t du H\intu(u;t_0)\NN\\
 \!&\!\times \!&\!\left[\un-i\int_{t_0}^u dv H\intu(v;t_0) \right] + \cdots ,
 \EEA
 where $H\intu(t;t_0)$, given by Eq.~(\ref{eq:HIMag}), contains a prefactor $\ep$.
 One then returns to the original representation with Eq.~(\ref{eq:UHIMag}).
 For the value of $A$ considered in Fig.~2, the Dyson expansion  yields the largest error $\Delta_1$ whereas its error on the transition probability is rather small.

In Fig.~3 we plot $g_2$ defined as the largest absolute value of the eigenvalues of the Hermitian matrix $G_2^{(2)}(t_f;t_i,t_1,t_1^\prime)$ given by Eq.~(\ref{eq:G2}).
This quantity which controls the error after one iteration is represented as a function of $t_1$ and $t_1^\prime$ for the KAM expansion of type B.
By minimizing $g_2$, which is the norm of this operator, with  respect to the free parameters  $t_1$ and $t_1^\prime$, one reduces the error  without having to determine the exact solution.
Notice that ($t_1=0,t_1^\prime=0$) is a local maximum of $g_2$ whereas 
($t_1=0.5,t_1^\prime=0.5$) is a saddle point.
The point ($t_1=0.5,t_1^\prime \approx 0.22$) corresponds to a minimum.
The symmetry of Fig.~3 results  from the pulse of Eq.~(\ref{pulse}) being symmetric.

Figure 4 displays the error $\Delta_1$ and the eigenvalue $g_2$ as a function of $t_1^\prime$ for the three types of one-iteration KAM expansions in the case $A=1$, $\ep=0.5$.
The value of $t_1$ is chosen so that the optimum can be reached: $t_1=0.5$ for the type B and $t_1=0.7$ for the type C (recall that the type A features no $t_1$).
One sees that the error can be reduced by more than one order of magnitude for the type B or C, and about half an order of magnitude for the type A after a single KAM iteration.
It is also seen that the eigenvalue $g_2$ is a very accurate estimation of the error $\Delta_1$, which enables one to locate the optimal values of the free parameters.
Note that the first order Magnus expansion corresponds to the  particular value $t_1^\prime=0$ (i.e. the nonoptimized case) of the one-iteration type A KAM expansion. 

We now turn to the next level of approximation for the  Magnus expansion and the  KAM expansion of type B.
Recall that the first order and one-iteration expansions of these schemes yield comparable errors $\Delta_1$ for  $A=1$.
Figure 5 shows that the (nonoptimized) two-iteration KAM expansion performs better than the second order Magnus expansion by one to two  orders of magnitude for $\Delta_2$.
The error $\delta_2$ on the transition probability is also much smaller for the KAM expansion.

It is worth noting from the comparisons of Figs.~2 and 5 that the error $\Delta_1$ for the optimized one-iteration KAM expansion is comparable to the error $\Delta_2$ for the nonoptimized two-iteration KAM expansion of type B. This conclusion is not restricted to the type B and can be understood on the basis of the Campbell-Baker-Hausdorff formula as discussed in Sec.~\ref{sec:comp}.

If one optimizes the type B KAM expansion by choosing $t_1=0.5$, $t_1^\prime=0.22 $ (as determined above from Fig.~3) and $t_2=0.66$, $t_2^\prime=0.8$, one gains another one to two orders of magnitude on $\Delta_2$.

From Fig.~5 one also deduces that the Dyson approach is not applicable in this context as the second order performs worse than the first order by both measures of the error.
Notice that the transition probability predicted by the Dyson technique diverges, as is well-known, by lack of unitarity.
In other words, the Dyson expansion does not allow one to refine the results of Fig.~2.

The two-iteration KAM expansion involves the operator $V_2(t)$ which is given by Eq.~(\ref{eq:V2KAM}) or (\ref{eq:V2KAMad}) as an infinite series of commutators.
For two-level systems, this series can be computed explicitly and results in Eq.~(\ref{eq:VnKAMresom}) with $n=1$.
It is remarkable that the coefficient $a_1$, $b_1$, $c_1$ and $d_1$ are well-defined for all values of $\ep$, even larger than unity.
In Fig.~5 we consider the cases where  $V_2(t)$ is truncated to two commutators [i. e. the term $k=1$ of order $\ep^0$ and the term $k=2$ of order $\ep$ in Eq.~(\ref{eq:V2KAMad})], and four commutators ($k=1,\cdots,4$).
Note that this amounts to approximate the coefficients of Eq.~(\ref{eq:abcd}) by polynomials (of order, respectively, 2 and 4) in $\ep \lambda_1(t)$.
The case of two commutators performs better than the second-order Magnus expansion which is not surprising as it contains all the terms of order $\ep^3$.
However, it performs worse than the KAM expansion with the infinite series by up to one order of magnitude. 
The case of four commutators is very close to the exact two-iteration KAM expansion for values of $\ep$ up to unity. The convergence with the number of commutators involved is indeed very fast. 
It has to be remarked that including commutators of higher orders in a well-defined manner as in the KAM algorithm is the main difference with the Magnus expansion (or any order by order expansion) and necessary to achieve superexponentiality.
From a practical point of view the inclusion of more than one commutator is both convenient and highly profitable.

The fact that for the KAM expansion the order of the remainder be the exponential of an exponential ($\epnm$) implies that the accuracy is increased in a dramatic way with the number $n$ of iterations. In addition, the accuracy of  a given $n$-iteration KAM expansion can still be enhanced, as illustrated here, and allows one to  get closer to the next iteration of the algorithm by including through the simple variation of the parameters $t_n$ and $t_n^\prime$ more appropriate higher order terms.

\begin{figure}[ht]
\includegraphics[scale=0.7]{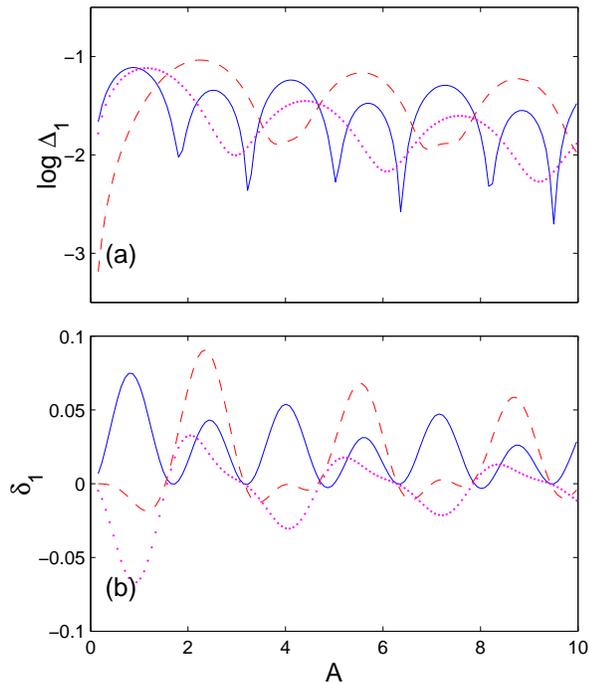}
\caption{\label{fig1}
Comparison of  the first order Magnus expansion (dots), the one-iteration type B KAM expansion (solid line) and the one-iteration type C KAM expansion (dashed line) for $\ep=0.5$ and $t_1=t_1^\prime=0$: (a) common logarithm of the error $\Delta_1$ and
(b) error $\delta_1$ as a function of $A$.
The one-iteration type A KAM expansion coincides in the case $t_1^\prime=0$ with the first order Magnus expansion. All quantities are dimensionless.}
\end{figure}

\begin{figure}[ht]
\includegraphics[scale=0.7]{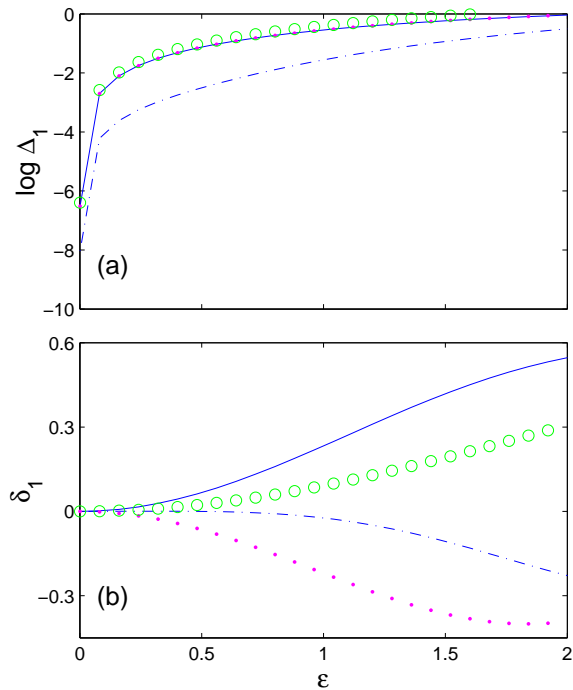}
\caption{\label{fig2}
Comparison of the first order Magnus expansion (dots), the first order Dyson expansion (circles), the one-iteration type B KAM expansion with $t_1=t_1^\prime=0$ (solid line) and the optimized one-iteration type B KAM expansion with $t_1=0.5$, $t_1^\prime=0.22$ (dash-dot line) for $A=1$: (a) common logarithm of the error $\Delta_1$  and
(b) error $\delta_1$ as a function of $\ep$. All quantities are dimensionless.}
\end{figure}

\begin{figure}[ht]
\includegraphics[scale=0.7]{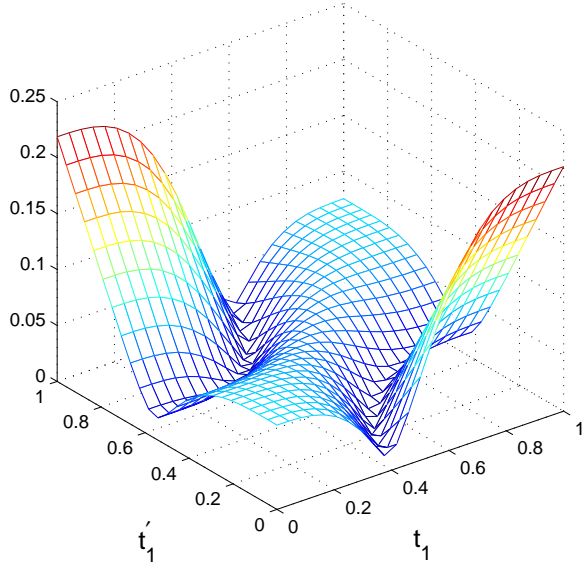}
\caption{\label{fig3}
Eigenvalue  $g_2$ for the KAM expansion of type B as a function of $t_1$ and $t_1^\prime$, for $A=1$ and $\ep=0.5$. All quantities are dimensionless.}
\end{figure}

\begin{figure}[ht]
\includegraphics[scale=0.7]{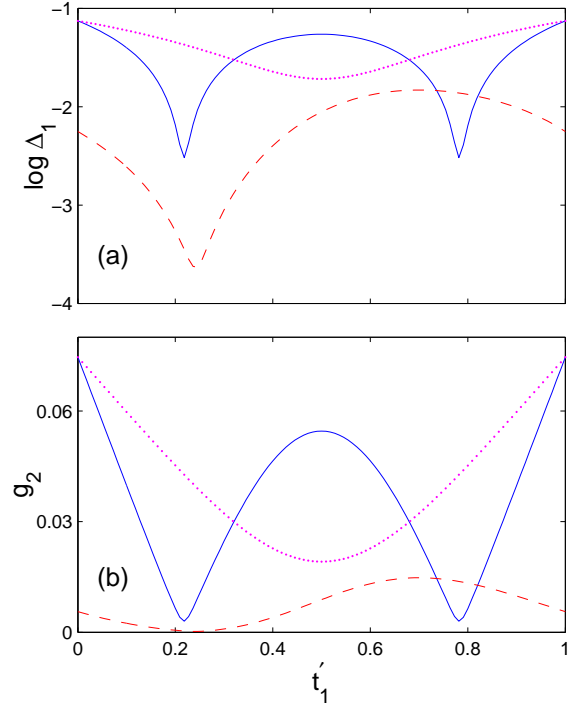}
\caption{\label{fig4}
Comparison of the one-iteration type A KAM expansion (dots), the one-iteration type B KAM expansion with $t_1=0.5$ (solid line) and the one-iteration type C KAM expansion with $t_1=0.7$ (dash-dot line) for $A=1$ and $\ep=0.5$: (a) common logarithm of the error $\Delta_1$  and (b) eigenvalue  $g_2$ as a function of $t_1^\prime$. All quantities are dimensionless.}
\end{figure}

\begin{figure}[ht]
\includegraphics[scale=0.7]{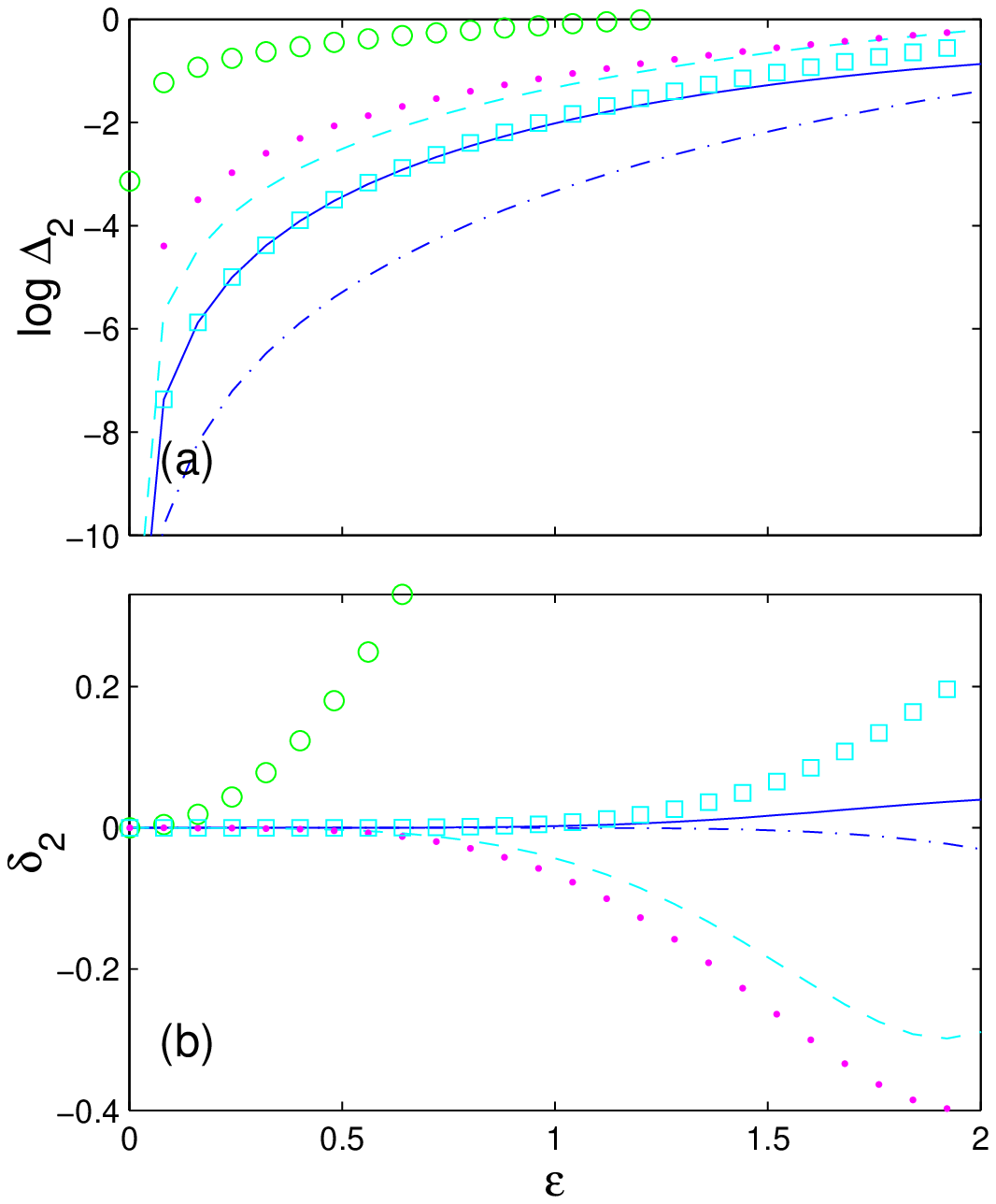}
\caption{\label{fig5}
Comparison of the second order Magnus expansion (dots), the second order Dyson expansion (circles), the two-iteration type B KAM expansion ($t_1=t_1^\prime=t_2=t_2^\prime=0$) with the infinite series of commutators (solid  line), two commutators (dashed line), four commutators (squares) and the optimized ($t_1=0.5$, $t_1^\prime=0.22$, $t_2=0.66$, $t_2^\prime=0.8$) two-iteration type B KAM expansion (dash-dot line)  for $A=1$: (a) common logarithm of the error $\Delta_2$  and
(b) error $\delta_1$ as a function of $\ep$. All quantities are dimensionless.}
\end{figure}

\section{Conclusions}
\label{sec:conclusion}
We have formulated perturbation theory in operator form \cite{Primas} for time-dependent problems localized in time directly in the original Hilbert space by unitarily transforming the evolution operator.

We have compared formally and numerically various unitary perturbative schemes.
The superiority of the KAM technique over the Magnus expansion, as well as the other methods, has been established   owing to its superexponential character and the accuracy optimization.
We have also shown that the Magnus expansion is recovered as a special case of the time-dependent Poincar\'e-Von Zeipel expansion (whose time-independent version coincides with the Rayleigh-Schr\"odinger expansion \cite{SchererI}). 
 
The  possibility to enhance the accuracy of a given level of approximation stems from  the free parameters and the free operators that appear naturally in the formulation presented here.
It allows one to significantly reduce  the error with respect to the exact solution without its knowledge by the minimization of an eigenvalue.

The above considerations, illustrated here on a pulse-driven two-level system, are straightforwardly applied to more involved problems (see Ref.~\cite{art3} for an application to the orientation and alignment of molecules).


\begin{acknowledgments}
The authors are grateful to G. Nicolis for helpful comments.
This research was supported in part by the Belgian FNRS, the {\it Action Concert\'{e}e
Incitative Photonique} from the French Ministry of Research, the
{\it Conseil R\'{e}gional de Bourgogne} and a CGRI-FNRS-CNRS cooperation.

\end{acknowledgments}

\appendix

\section{Time-dependent Poincar\'e-Von Zeipel expansion}
\label{app:PV}
The time-dependent  Poincar\'e-Von Zeipel technique amounts to construct a unitary operator $T(t)$ which transforms  the propagator $U_{H_1}(t,t_{0})$
into a propagator $U_{H\ef}(t,t_0)$ according to
\BEA
\label{eq:TUHPV}
T^{\dagger}(t)U_{H_1}(t,t_0)T(t_0)=U_{H\ef}(t,t_0) ,
\EEA
where $U_{H\ef}(t,t_{0})$ is associated with an effective Hamiltonian $H\ef(t)$ containing contributions to every order in $\ep$
\BE
\label{eq:UHedotPV}
H\ef(t)\equiv H_0(t)+\sum_{k=1}^{\infty}\ep^k D_k(t;t_k) .\quad
\EE
This propagator 
can be expressed in terms of $U_{H_0}(t,t_{0})$ and unitary operators related to the partial Hamiltonians $\ep^k D_k(t;t_k)$
\BEA
U_{H\ef}(t,t_{0})\!&\!= \!&\!
U_{H_0}(t,t_0)  \exp \left[-i (t-t_0) \ep D_{1}(t_0;t_1)\right]\NN\\
 \!&\! \times  \!&\!\cdots    \exp \left[-i (t-t_0) \ep^k D_{k}(t_0;t_k)\right] \cdots, \qquad
\label{eq:UHePV}
\EEA
provided the operator $D_k(t;t_k)$ are defined as
\BEA
\label{eq:DnPV}
D_k(t;t_k)\equiv U_{H\efkm}(t,t_k)D_k(t_k;t_k)U_{H\efkm}(t_k,t).\quad
\EEA
Here $D_k(t_k;t_k)$ is arbitrary but strictly of order $\ep ^0$ (in contrast to the KAM case) and we set
\BEA
U_{H\efn}(t,t_0)\!&\!\equiv \!&\!
U_{H_0}(t,t_0)  \exp \left[-i (t-t_0) \ep D_{1}(t_0;t_1)\right]\NN\\
 \!&\!\times   \!&\! \cdots   \exp \left[-i (t-t_0) \ep^n D_{n}(t_0;t_n)\right]  . \quad
\label{eq:UHefnPV}
\EEA
The generator of the Poincar\'e-Von Zeipel transformation $T(t)$ is written as a power series of $\ep$-independent operators $W_k(t)$
\BEA
\label{eq:TPV}
T(t)\equiv \exp\left(-i\sum_{k=1}^{\infty}\ep^k W_k(t)\right) .
\EEA
These operators satisfy the  differential equations
\BEA
\label{eq:WndotPV}
\frac{\partial}{\partial t}W_k(t)=V_k(t)-D_k(t;t_k)+i\left[W_k(t),H_0(t)\right] ,
\EEA
where the expression for $V_k(t)$  cannot be given in a simple form for arbitrary $k$ and requires increasing algebra. 
The situation is analogous for the Magnus expansion or the Van Vleck expansion given below, and in contrast to the KAM expansion where the new effective perturbation has exactly the same form at each step [cf. Eq.~(\ref{eq:VnKAM})].   
The general solution to Eq.~(\ref{eq:WndotPV})  reads (up to a term $U_{H_0}(t,t_0)B_k U_{H_0}(t_0,t)$ with $B_k$ any constant self-adjoint operator)
\BEA
\label{eq:WnPV}
W_k(t)=\int_{t_k^{\prime}}^t d u U_{H_0}(t,u)\left[V_k(u)-D_k(u)\right]U_{H_0}(u,t), \quad
\EEA
where $t_k^{\prime}$ is arbitrary.
Notice that this expression involves the unperturbed propagator unlike the KAM analogue which features an effective propagator [cf. Eq.~(\ref{eq:WnKAM})].

For the operator $D_k(t;t_k)$ of Eq.~(\ref{eq:DnPV}) to be strictly  of order $\ep^0$, we are no longer entitled to choose $U_{H\efkm}(t,t_k) V_k(t_k) U_{H\efkm}(t_k,t)$ for any $k$  if that choice has already been made for a lower value of $k$.
This choice can only be made once (for an arbitrary value of $k$ denoted $v$) and implies to take 
$D_k(t;t_k)\equiv 0$ for the other values of $k$.

We then construct the $n$-th order unitary approximation to $T(t)$
\BEA
\label{eq:TnPV}
T_n(t)\equiv \exp\left(-i\sum_{k=1}^{n}\ep^k W_k(t)\right), 
\EEA
and by Eqs.~(\ref{eq:TUHPV}) and (\ref{eq:UHePV}) obtain the $n$-th order Poincar\'e-Von Zeipel approximation to the propagator $U_{H_1}(t,t_0)$
\BEA
\label{eq:UHnPV}
U_{H_1}^{(n)}(t,t_0)\!&\!=\!&\!T_n(t)U_{H_0}(t,t_0) \NN\\
\!&\! \times \!&\! \exp \left[-i (t-t_0) \ep^v D_{v}(t_0;t_v)\right] T_n^{\dagger}(t_0) ,\qquad
\EEA
where $v$ is any integer between 1 and $n$ for which we have the possibility to choose either $D_{v}(t_0;t_v) \equiv U_{H_0}(t_0,t_v) V_{v}(t_v) U_{H_0}(t_v,t_0)$ or $D_{v}(t_0;t_v)\equiv 0$.

It is interesting to note that Eq.~(\ref{eq:UHnPV}) precisely  reduces to the Magnus expansion if we take $D_v(t;t_v)\equiv 0$ and $t_k^{\prime}=t_0$ for all $k$.
In the general case there are up to $n+1$ free parameters
\BEA
T_n(t)\!&\!\equiv \!&\!T_n(t;t_v,t_1^\prime,\cdots,t_n^\prime),\NN\\
W_n(t)\!&\!\equiv \!&\!W_n(t;t_v,t_1^\prime,\cdots,t_n^\prime).
\EEA

\section{Time-dependent Van Vleck method}
\label{app:VV}
The time-dependent Van Vleck technique is an order by order method where instead of transforming the original propagator into a final new propagator in a single step as in the Poincar\'e-Von Zeipel algorithm, one achieves this goal iteratively through a series of transformations $T_k(t)$ which reduce the size of correction terms from $\ep^k$ to $\ep^{k+1}$.
This results in the $n$-th order Van Vleck expansion
\BEA
\label{eq:UHT2VV}
U_{H_1}^{(n)}\!\!& \!\!&\!\!(t,t_0) = {T_1}(t)\cdots{T_n}(t)U_{H_0}(t,t_0) \NN\\
\times\!\!&\!\! &\!\!\exp \left[-i (t-t_0) \ep^v D_{v}(t_0;t_v)\right] T_n^{\dagger}(t_0)\cdots T_1^{\dagger}(t_0),\qquad
\EEA
where $D_{v}(t_0;t_v)\equiv U_{H_0}(t_0,t_v) V_{v}(t_v) U_{H_0}(t_v,t_0)$ or $D_{v}(t_0;t_v)\equiv 0$, with $v$  an integer between 1 and $n$.
Here we define
\BE
T_k(t)\equiv e^{-i \ep^k W_k(t)},
\EE
with $W_k(t)$ constructed as in Eq.~(\ref{eq:WnPV}).
There are up to $n+1$ free parameters entering the $n$-th order time-dependent Van Vleck expansion.

The Van Vleck  and  Poincar\'e-Von Zeipel  techniques differ   because of the product of exponentials of $W_k(t)$ appearing in Eq.~(\ref{eq:UHT2VV}) instead of the single exponential of a sum of $W_k(t)$ in Eq.~(\ref{eq:UHnPV}). 
As discussed above on the basis of the Campbell-Baker-Hausdorff formula  this means that these algorithms differ at orders higher than the prescribed order.

\section{KAM expansion in the interaction representation}
\label{app:KAM}
In the framework of a perturbation theory, we stressed in Sec.~\ref{sec:Mag} that the Magnus expansion had to be derived in the interaction representation.
In the section above,  the KAM algorithm was applied in the original representation.
We show here that going to the interaction representation, applying the KAM technique and coming back to the original representation yields identically the same expansions as above despite the truncation at any finite order.

Given an Hamiltonian $H_1(t)=H_0(t)+\ep V_1(t)$ and the propagator $U_{H_0}(t,t_0)$ we consider the interaction representation with respect to $H_0(t)$.
From Eq.~(\ref{eq:interaction}) one deduces that $H\intu(t;s) = H\intz(t;s) + \ep  V\intu(t;s)$ with
\BES
\BEA
 H\intz(t;s) \!\!&\equiv & 0,\\
 V\intu(t;s)&\equiv&\!\! U_{H_0}(s,t)  V_1(t)U_{H_0}(t,s)    .
 \label{eq:H1i}
\EEA
Furthermore, the unperturbed propagator is trivial
\BE
\label{eq:UH0i}
U_{H\intz}(t,t_0;s)=\un .\qquad
\EE
\EES
It is worth pointing out that Eq.~(\ref{eq:H1i}) suggests that $\ep  V\intu(t;s) $ is considered small with respect to $H\intz(t;s)=0$.
As a matter of fact, by virtue of Eq.~(\ref{eq:XIdot}), it is with respect to $H\intz(t;s)-i\ddpt$ that $\ep  V\intu(t;s) $ is considered small (in a technical sense we need not specify here).

Applying the KAM algorithm with the identifications of Eqs.~(\ref{eq:H1i}) and (\ref{eq:UH0i}) leads to the  expansion
\BEA
U_{H\intu}^{(n)}(t,t_0;s)\!\!& = &\!\!{T_1}(t;s)\cdots T_{n}(t;s)U_{H\eintn}(t,t_0;s)\NN\\
&\times&\!\!T_{n}^{\dagger}(t_0;s)\ldots T_1^{\dagger}(t_0;s). \qquad \quad
\label{eq:UH1in}
\EEA
 
Returning to the original representation with the help of Eq.~(\ref{eq:UHIMag}) one obtains a perturbative expansion for $U_{H_1}(t,t_0)$ that coincides exactly with the expansion $U_{H_1}^{(n)}(t,t_0)$ obtained directly in this representation
\BE
 U_{H_0}(t,s) U_{H\intu}^{(n)}(t,t_0;s) U_{H_0}(s,t_0)=U_{H_1}^{(n)}(t,t_0).
\EE
This  stems from the following identities readily derived on the basis of  Eq.~(\ref{eq:idexp}) and valid for any $k$
\BES
\BEA
U_{H_0}(t,s) T_{k}(t;s) U_{H_0}(s,t) &=&T_{k}(t) ,\\
U_{H_0}(t,s) U_{H\eintk}(t,t_0;s) U_{H_0}(s,t_0) &=&U_{H\efk}(t,t_0).\qquad
\EEA
\EES

\end{document}